\documentclass[]{aastex}
\usepackage{emulateapj5,epsfig}

\newlength{\colwidth}
\newcommand{\hMsun}{{h^{-1}}{\rm M}_{\solar}}
\newcommand{\hmsun}{{h^{-1}}{\rm M}_{\solar}}
\newcommand{\solar}{_{\mathord\odot}}
\newcommand{\hmpc}{\ifmmode{h^{-1}{\rm Mpc}}\;\else${h^{-1}}${\rm Mpc}\fi}
\newcommand{\hkpc}{\ifmmode{h^{-1}{\rm kpc}}\;\else${h^{-1}}${\rm kpc}\fi}

\newcommand{\LCDM}{$\Lambda$CDM}
\newcommand{\beq}{\begin{equation}}
\newcommand{\eeq}{\end{equation}}

\newenvironment{inlinefigure}{
\def\@captype{figure}
\noindent\begin{minipage}{0.999\linewidth}\begin{center}}
{\end{center}\end{minipage}\smallskip}

\def\rsp{r_{\rm sp}}
\def\rspo{r_{\rm sp1}}
\def\rspt{r_{\rm sp2}}
\def\Np{N_{\rm p}}
\def\Npmin{N_{\rm p}^{\rm min}}
\def\Npfmin{N_{\rm p,fit}^{\rm min}}
\def\Mvirmin{M_{\rm vir}^{\rm min}}

\def\Rvirmin{R_{\rm vir}^{\rm min}}
\def\Nspo{N_{\rm sp1}}
\def\Nspt{N_{\rm sp2}}
\def\ni{\noindent}
\def\lsim{\lower.5ex\hbox{\ltsima}}
\def\ltsima{$\; \buildrel < \over \sim \;$}

\newcommand{\cv}{\ifmmode{c_{\rm vir}}\else$c_{\rm vir}$\fi}
\newcommand{\Cv}{\ifmmode{c_{\rm vir}}\else$c_{\rm vir}$\fi}
\newcommand{\cvir}{\ifmmode{c_{\rm vir}}\else$c_{\rm vir}$\fi}
\newcommand{\Cvir}{\ifmmode{c_{\rm vir}}\else$c_{\rm vir}$\fi}
\newcommand{\Rs}{\ifmmode{R_{\rm s}}\else$R_{\rm s}$\fi}
\newcommand{\rs}{\ifmmode{R_{\rm s}}\else$R_{\rm s}$\fi}
\newcommand{\Rt}{\ifmmode{R_{\rm t}}\else$R_{\rm t}$\fi}
\newcommand{\rt}{\ifmmode{R_{\rm t}}\else$R_{\rm t}$\fi}
\newcommand{\Rvir}{\ifmmode{R_{\rm vir}}\else$R_{\rm vir}$\fi}
\newcommand{\rvir}{\ifmmode{R_{\rm vir}}\else$R_{\rm vir}$\fi}
\newcommand{\Mvir}{\ifmmode{M_{\rm vir}}\else$M_{\rm vir}$\fi}
\newcommand{\mvir}{\ifmmode{M_{\rm vir}}\else$M_{\rm vir}$\fi}
\newcommand{\mt}{\ifmmode{M_{\rm t}}\else$M_{\rm t}$\fi}
\newcommand{\Mt}{\ifmmode{M_{\rm t}}\else$M_{\rm t}$\fi}
\newcommand{\Mh}{\ifmmode{M_{\rm h}}\else$M_{\rm h}$\fi}
\newcommand{\Rh}{\ifmmode{R_{\rm h}}\else$R_{\rm h}$\fi}
\newcommand{\Mnfw}{\ifmmode{M_{\sss \rm NFW}}\else$M_{\sss \rm NFW}$\fi}
\newcommand{\Mtraj}{\ifmmode{M_{\rm traj}}\else$M_{\rm traj}$\fi}
\newcommand{\ac}{\ifmmode{a_{\rm c}}\else$a_{\rm c}$\fi}
\newcommand{\act}{\tilde{a}_{\rm c}}
\newcommand{\cnfw}{\ifmmode{c_{\sss \rm NFW}}\else$c_{\sss \rm NFW}$\fi}
\newcommand{\Ms}{\ifmmode{M_{*}}\else$M_{*}$\fi}
\newcommand{\mstar}{\ifmmode{M_{*}}\else$M_{*}$\fi}
\newcommand{\rhonfw}{\rho_{\sss \rm NFW}}
\newcommand{\rhos}{\rho_{\rm s}}
\newcommand{\rhou}{\rho_{\rm u}}
\newcommand{\vvir}{V_{\rm vir}}
\newcommand{\vmax}{V_{\rm max}}
\newcommand{\omm}{\Omega_{\rm m}}
\newcommand{\oml}{\Omega_{\Lambda}}
\newcommand{\sss}{\scriptscriptstyle}
\newcommand{\fres}{f_{\rm res}}
\newcommand{\Dvir}{\Delta_{\rm vir}}

\newcommand{\lamf}{\ifmmode{\lambda_f}\;\else${\lambda_f}$\fi}

\newcommand{\cc}{\ifmmode{c_{\rm 1}}\else$c_{\rm 1}$\fi}
\newcommand{\zf}{\ifmmode{z_{\rm f}}\else$z_{\rm f}$\fi}
\newcommand{\zc}{\ifmmode{z_{\rm c}}\else$z_{\rm c}$\fi}
\newcommand{\zobs}{\ifmmode{z_{\rm o}}\else$z_{\rm o}$\fi}
\newcommand{\aobs}{\ifmmode{a_{\rm o}}\else$a_{\rm o}$\fi}
\newcommand{\Mobs}{\ifmmode{M_{\rm o}}\else$M_{\rm o}$\fi}

\shortauthors{WECHSLER et al.}
\shorttitle{CONCENTRATION AND HALO ASSEMBLY HISTORY}
\begin{document}
\submitted{Submitted 2001 August 9; last revision 2001 October 26}
\journalinfo{The Astrophysical Journal, in press}

\title{Concentrations of Dark Halos from their Assembly Histories}

\author{Risa H. Wechsler\altaffilmark{1}, 
         James S. Bullock\altaffilmark{2}, 
        Joel R. Primack\altaffilmark{1}, 
	Andrey V. Kravtsov\altaffilmark{2,3}, 
        Avishai Dekel\altaffilmark{4},
}

\altaffiltext{1}{Physics Department, University of California, Santa Cruz, 
        CA 95064; wechsler@umich.edu, joel@ucolick.org}
\altaffiltext{2}{Department of Astronomy, The Ohio State 
        University, 140 W. 18th Ave, Columbus, OH 43210;
        james@astronomy.ohio-state.edu, andrey@oddjob.uchicago.edu}
\altaffiltext{3}{Hubble Fellow}
\altaffiltext{4}{Racah Institute of Physics, The Hebrew 
        University, Jerusalem 91904 Israel; dekel@astro.huji.ac.il}

\begin{abstract}
We study the relation between the density profiles of dark matter
halos and their mass assembly histories, using a statistical sample of
halos in a high-resolution N-body simulation of the \LCDM\ cosmology.
For each halo at $z=0$, we identify its merger-history tree, and
determine concentration parameters $\cvir$ for all progenitors, thus
providing a structural merger tree for each halo.  We fit the mass
accretion histories by a universal function with one parameter, the
formation epoch \ac, defined when the log mass accretion rate ${\rm
d}\log M/{\rm d}\log a$ falls below a critical value $S$.  We find
that late forming galaxies tend to be less concentrated, such that
$\cvir$ ``observed'' at any epoch $\aobs$ is strongly correlated with
$\ac$ via $\cvir=\cc \aobs /\ac$.  Scatter about this relation is
mostly due to measurement errors in $\cvir$ and $\ac$, implying that
the actual spread in $\cvir$ for halos of a given mass can be mostly
attributed to scatter in $\ac$.  We demonstrate that this relation can
also be used to predict the mass and redshift dependence of \cv, and
the scatter about the median $\cv(M,z)$, using accretion histories
derived from the Extended Press-Schechter (EPS) formalism, after
adjusting for a constant offset between the formation times as
predicted by EPS and as measured in the simulations; this new
ingredient can thus be easily incorporated into semi-analytic models
of galaxy formation.  The correlation found between halo concentration
and mass accretion rate suggests a physical interpretation: for high
mass infall rates the central density is related to the background
density; when the mass infall rate slows, the central density stays
approximately constant and the halo concentration just grows as \Rvir.
Because of the direct connection between halo concentration and
velocity rotation curves, and because of probable connections between
halo mass assembly history and star formation history, the tight
correlation between these properties provides an essential new
ingredient for galaxy formation modeling.
\end{abstract}


\section{INTRODUCTION}
\label{sec:intro}
The theory of cold dark matter (CDM) \citep[e.g.,][]{peebles:82,
blumen:84, davis:85} provides a remarkably successful framework for
understanding galaxy assembly and structure formation in the universe.
Within this picture, dark matter collapses first into small halos, and
these halos merge to form progressively larger halos over time.  As
this process continues, baryons that initially trace the dark matter
cool and condense to form galaxies in halo centers.  New supplies of
gas and galaxy mergers are closely linked to the mass accretion
histories of the halos they inhabit.  A detailed understanding of how
this mass accretion occurs, and how individual halo properties depend
on their merger histories, is of fundamental importance for predicting
galaxy properties within the CDM theory, and, similarly, for using
observed galaxy properties (e.g., rotation curves) to test the
paradigm.

The basic theory for the buildup of structure in the universe, and for
the evolution of the properties of gravitationally-bound structures,
is fairly well developed; it has been extensively simulated at
increasingly high resolution, and analytic formulations have been
developed to describe this behavior.  The Press-Schechter formalism
\citep{ps:74,bond:91} has provided a useful framework for
understanding the mass function of dark halos, and, with subsequent
improvements, has been shown to work well in comparison to N-body
simulations (\citealt{gross:98, st:99}; \citealt*{smt:01};
\citealt{jenk:01,st:01}).  The theory has been extended using an
excursion-set formalism to an Extended Press-Schechter theory (EPS),
which describes the statistics of the buildup of individual halos
through time \citep{bond:91,lacey:93}.  This EPS theory has been
implemented to construct random realizations of merger trees, each
specifying a full assembly history for a halo
\citep[e.g.,][hereafter SK99]{lacey:93, kauf:93, sk:99}.
Detailed comparisons of EPS with simulations have highlighted the general
success of the theory and quantified the level of accuracy of its
predictions (e.g., \citealt{somerville:00}, hereafter SLKD; \citealt*{gardner:00, cohn:01}).

Similar advances have been made in understanding halo density
structure \citep{ef:88,frenk:88,dc:91}.  Navarro, Frenk, \& White
(\citeyear{nfw:95},\citeyear{nfw:96},\citeyear{nfw:97}; hereafter
NFW), have proposed that halo profiles can be universally fit by a
two-parameter functional form:
\begin{equation}
\rhonfw(r) = \frac{\rhos}{(r/\rs)\left(1+r/\rs\right)^2},
\label{eq:nfw}
\end{equation}
where $\rs$ is a characteristic ``inner'' radius, and $\rhos$ a
corresponding inner density. One of the inner parameters can be
replaced by a ``virial'' parameter, either the virial radius
($\rvir$), mass ($\Mvir$), or velocity ($\vvir$), defined such that
the mean density inside the virial radius is $\Dvir$ times the mean
universal density $\rho_u$ at that redshift:
\beq
\Mvir  \equiv  \frac{4 \pi}{3} \Dvir \rho_u \rvir^3.
\label{eqt:mvir}
\eeq
The critical overdensity at virialization, $\Dvir$, is motivated by the
spherical collapse model; it has a value $\simeq 180$ for the
Einstein-deSitter cosmology, and $\simeq 340$ for the $\Lambda$CDM
cosmology assumed here.  A useful alternative parameter for describing
the shape of the profile is the concentration parameter $\cvir$,
defined as $\cvir\equiv\rvir/\rs.$ NFW found that this functional form
provides a good fit to halos over two decades in radius, for a large
range of halo masses, and for several different cosmological
scenarios.  They tested it for the Einstein-deSitter model with a
standard CDM power spectrum of initial fluctuations (SCDM), a flat
cosmological model with $\omm=0.3$, $\oml=0.7$ and a corresponding CDM
power spectrum ($\Lambda$CDM), and several models with power-law power
spectra (confirmed by \citealt{craig:97}, and \citealt*{kkk:97}).

Later studies (e.g., \citealt{moore:98, ghigna:00,kkbp:01}) have
indicated that the inner logarithmic slopes of at least some halo
density profiles in CDM cosmologies are closer to $-1.5$ than to
$-1.0$.~\footnote[1]{The opposite suggestion has also been made: that
the asymptotic slope as $r\rightarrow0$ may be somewhat shallower than
$-1.0$ \citep{tn:01}.}  However, \citet{kkbp:01} showed that even for
halos that are better fit by $-1.5$ slopes, an NFW fit is perfectly
adequate outside the inner $\sim1$\% of the virial radius (this
corresponds to roughly $\sim 3$kpc for a Milky-Way type halo).  An
important implication (as pointed out by \citealt{bullock:00},
hereafter B01) is that even if halos are not perfectly described by
the NFW form (see also \citealt{av:99} and \citealt{jing:00}), fit
parameters derived using this profile provide a useful general
characterization of the density structure, relating, for example, the
central density of a halo to that of the background via two fit
parameters (e.g., $\cvir$ and $\Rvir$).

The two-parameter NFW characterization of halo profiles has provided
several useful insights into the nature of halo collapse.  Among the
most interesting results, first noticed by NFW, was that, for a given
cosmology, the central density $\rhos$ varies inversely with halo
mass, or, equivalently, $\cvir$ tends to increase with decreasing
$\Mvir$.  A natural reason for this fact is that low-mass halos
typically collapse earlier, when the universe was denser, and the
central density somehow reflects this higher background
density.~\footnote[2]{ Indeed, for models with truncated power
spectra, in which there is no systematic relation between collapse
time and halo mass, no correlation is seen between mass and $\cvir$
(\citealt{av:01}; \citealt*{bode:01,ens:01}).}  In a toy model to
explain this correlation, NFW (1997) assumed that $\rhos$ is a
constant multiple $k$ of the universal density $\rhou(\zc)$ at a
collapse redshift $\zc$. They defined the collapse redshift as the
time when half of the halo's mass was first in progenitors more
massive than $f$ times the halo's mass.  The general trend of the
relation between the two profile parameters at $z=0$ is reproduced
well using EPS to predict $\zc$, and a proper choice of values for the
constants $k$ and $f$ (favored parameters for \LCDM\ were $f=0.01$ and
$k=3.4\times{10}^{3}$).

Subsequent analyses revealed additional complexities and trends in
halo structure.  First, it has been realized that while the
$\Mvir$-$\cvir$ trend holds on average, halos of fixed mass show
significant scatter in their $\cvir$ values (\citealt{bullock:99th,
jing:00}; B01). In the context of the proposed correlation between
concentration and collapse time, the scatter in $\cvir$ is a natural
reflection of the expected scatter in collapse time for halos of a
given mass.

B01 also found that halo concentrations (at fixed mass) are
systematically lower at higher redshifts, $\cvir \propto 1/(1+z)$,
implying a much stronger trend than that predicted by the NFW toy
model.  Since by definition the concentration at redshift $z$ roughly
relates the halo central density to the universal background density
at that time via $\cvir(z) \propto [\rho_c(z)/\bar{\rho}(z)]^{1/3} $,
we have $\cvir(z) \propto \rho_c(z)^{1/3}/(1+z)$.  The observed
relation implies that a halo's central density (or collapse time) is
set only by the halo's mass, {\textit{independent of the redshift when
the halo is observed}}.  One reason for the shortcoming of the NFW toy
model in reproducing the proper redshift dependence is that the
collapse time as defined by this model also depends on the redshift of
observation. In order to properly match the time evolution, B01
proposed a modified toy model in which the collapse time, denoted now
by $\act$ (instead of $\ac$ to avoid later confusion) is set by the
mass only, and is defined as the epoch at which the typical collapsing
mass, $\Ms(\act)$, equals a fixed fraction $F$ of the halo mass at
epoch $a$, $\Ms(\act) \equiv F \Mvir$.  The typical collapsing mass is
defined by $\sigma [\Ms(a)]=1.686/D(a)$, where $\sigma (M)$ is the rms
density fluctuation on the comoving scale encompassing a mass $M$,
extrapolated using linear theory to the present, $a=1$, and $D(a)$ is
the linear growth rate.  The implied concentration is given by
$\cvir(a) = K a/\act$, which, with appropriate values of $F$ and $K$,
reproduces quite well the dependence of concentration on both mass and
redshift as measured in simulations (see B01 for details).

Simplified models of the type discussed so far provide a qualitative
understanding and useful parameterization of the complex processes
seen in simulations (for further insights and an exploration of
non-hierarchical power spectra see \citealt{ens:01}).  However, our
understanding remains somewhat schematic and several important
questions remain open.  First, how do {\em individual} halo density
profiles build up over time?  How do the mass accretion histories
affect the final halo concentrations, and how can physically realistic
mass accretion histories be connected to the simplified definition of
formation time advocated by B01?  Can EPS be used to predict halo
concentrations?  What physical process is responsible for the scatter
in $\cvir$ at fixed mass?  This work builds on the work of B01, using
the same simulations analyzed there, together with a new ``structural
merger tree'' described below.  The goal is to see if we can directly
correlate the assembly history of halos with their dark matter halo
density profiles, test the model proposed by B01, and develop a
physical model for the range of halo concentrations seen in
simulations, including scatter and dependence on mass and redshift.

These questions are interesting from a theoretical perspective, but
they also have direct observational implications.  The shape of the
halo density profile directly affects the rotation curve of the galaxy
that forms within it; the one-sigma scatter in concentration observed
by B01 corresponds to, for a $1\times10^{11}\hmsun$ halo, a scatter
in $\vmax$ values of $\sim 85-105$ km/s.  Thus scatter in halo
density profiles is directly related to scatter in the Tully-Fisher
relation, and also has implications for the relative contributions of
halo and disk to velocity rotation curves.  In addition, density
profiles may affect other aspects of the galaxy formation process,
such as the efficiency of gas cooling or the star formation rate.  If
halo concentrations are related to their mass assembly histories, it
may indicate that halo concentration is correlated with galaxy type.
This would have implications for both the zero-point and scatter in
the Tully-Fisher relation, and could also have implications for the
rotation curves of low-surface brightness galaxies.  Possible
correlations between halo density profiles and galaxy observables
(with the exception of mass) have been neglected in previous modeling
efforts, but it is clear that such correlations are likely to be quite
important.

We begin in \S \ref{sec:halos}  by summarizing the relevant details of
the N-body simulation, halo finder and density profile fitting.  In \S
\ref{sec:tree}, the ``structural merger tree'' developed for this
project is described.  We continue in \S \ref{sec:mah} by detailing
how mass accretion histories are constructed and then describe a new
method for defining a characteristic formation epoch for each halo. We
then show how this formation epoch can be related to the halo
concentration, and how this can explain the dependence of
concentration on mass and redshift as well as explaining the origin
and magnitude of the scatter in these relations.  In \S
\ref{sec:eps}, we show how Extended Press-Schechter theory can be used
to predict concentrations for individual halos using our model for
relating halo concentration to characteristic formation epoch.  This
model can reproduce the scatter, mass and redshift trends observed in
N-body simulations.  In \S \ref{sec:scatter}, we discuss the scatter
in \cvir(\Mvir), and how it depends on the merging history of halos.
We discuss the implications of our results and conclude in \S
\ref{sec:c_conclude}.

\section{SIMULATED HALOS}
\label{sec:halos}
In the work that follows, we consider only one cosmology whose
evolution has been simulated with the ART code \citep*{kkk:97}, a flat
\LCDM\ model with $\Omega_m = 0.3$, $h = 0.7$ and $\sigma_8$ = 1.0.
The simulation is the same as that used in B01.  It followed the
trajectories of $256^3$ cold dark matter particles within a cubic,
periodic box of comoving size 60\hmpc\ from redshift $z = 40$ to the
present.  A $512^3$ uniform grid is used, with up to six
refinement levels in the regions of highest density, implying a
dynamic range of $32,768$.  The formal resolution of the simulation is
thus $\fres = 1.8\hkpc$, and the mass per particle is $m_{p} = 1.1
\times 10^{9} \hMsun.$  
For the purpose of constructing accurate merger trees, here we analyze
the simulation data from 36 output times thinly spaced between $z=7$
and $0$.  It should be noted that the 

\begin{inlinefigure} 
\centerline{\epsfxsize=\colwidth\epsffile{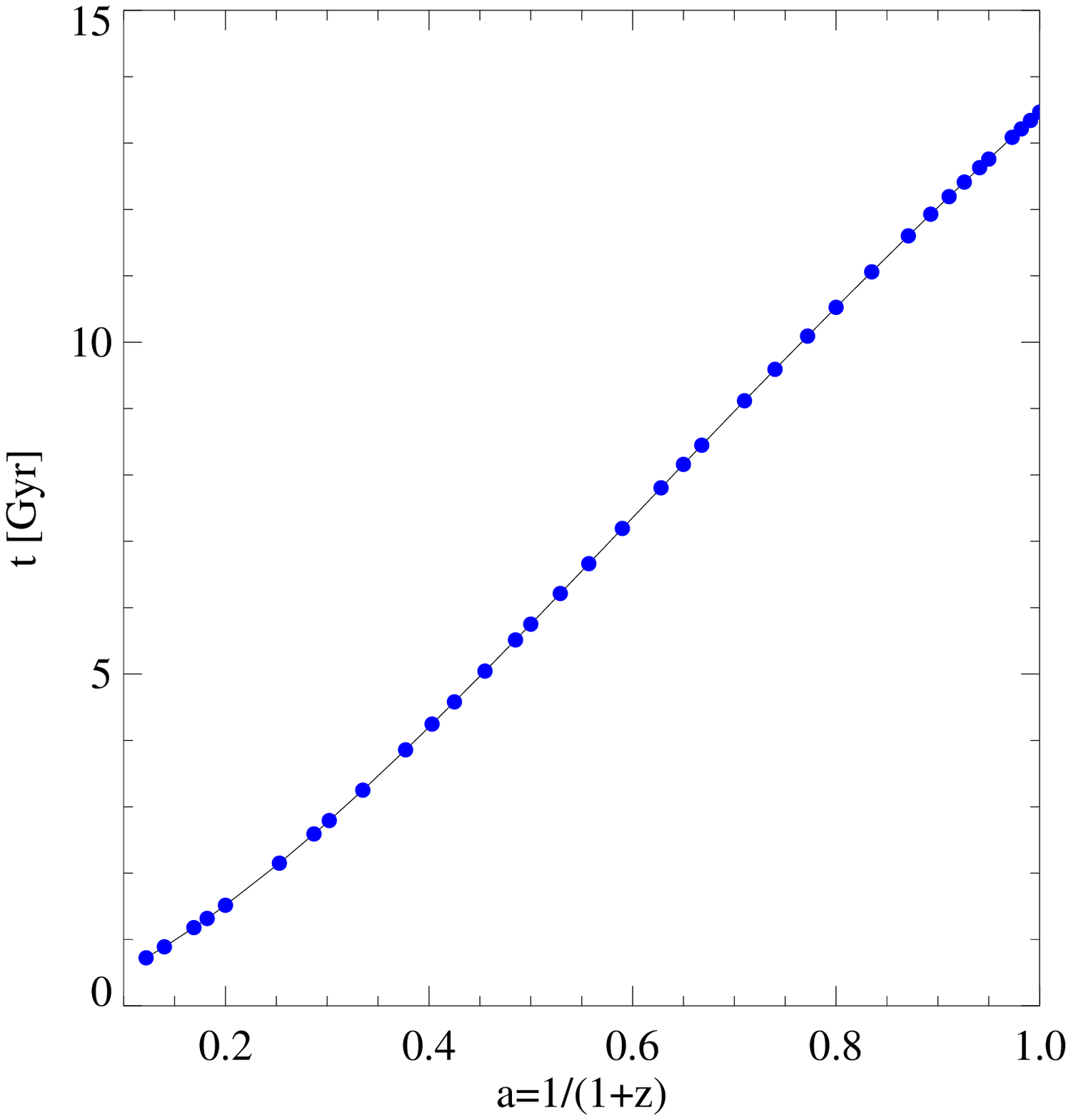}}
\figcaption{
Output times of the simulation. The curve is the age of the universe
$t$ as a function of the universal expansion factor $a$ for the \LCDM\
model considered here.  The symbols mark the 36 output times of the
simulation.
\label{fig:time}}
\end{inlinefigure}

\noindent methods described here for
finding and fitting halos and constructing merger trees are completely
generalizable to other simulations or cosmologies.

The halo finding algorithm used here is based on the Bound Density
Maxima \citep{kh:97} technique described in \citet{bullock:99th} and
B01, but we have modified and optimized it for the purpose of building
a structural merger tree.  The essential elements are presented
briefly here, and details are described in the Appendix.  For each
density maximum, we step out in radial shells until the mean
overdensity falls below $\Delta_{\rm vir}(z)$, \nolinebreak[3] 
\footnote[3]{For flat cosmologies,  $\Dvir$ can be approximated
by \citep{bryan:98} $\Dvir \simeq (18\pi^2 + 82x - 39x^2)/\Omega(z)$,
where $x\equiv \Omega(z) - 1$, and $\Omega(z)$ is the ratio of mean
matter density to critical density at redshift $z$; for the \LCDM\
model considered here, $\Dvir\simeq 337$.}  or the radial profile
shows a significant upturn.  We denote this radius as $\Rh$ and the
mass determined by counting particles within this radius as $\Mh$.  We
attempt to identify halos containing as few as $\Npmin=20$ particles;
our halo catalog thus includes $\sim 14000$ halos above a mass
threshold of $2.2\times10^{10}\hMsun$.  By comparing our simulation
with a smaller, higher resolution realization of the same cosmology,
at this mass we estimate our completeness to be $70$\%, and we are
$\sim100\%$ complete at $6.6\times10^{10}\hMsun$.  NFW profiles are
fit to halos with more than $\Npfmin=200$ particles, corresponding to
halos more massive than $2.2\times10^{11}\hMsun$.  A profile fit to a
halo of only a few hundred particles carries large errors, but as long
as the fit is done properly to include these errors, reliable
concentrations estimates can be obtained (B01).  Halos in the mass
range $2\times10^{11} - 5\times10^{11} \hMsun$ have typical fit mass
errors of about 10\%, and $\cvir$ errors of 15-20\%.  However, a few
percent of the time the errors are significantly larger than that.  We
therefore make a rigorous attempt to estimate the errors and take them
into account in every step of the process.  Poor fits are marked by
large errors that are incorporated in the analysis and the results we
present.  The outcome at any output time is a statistical halo catalog
that includes all the bound virialized systems in the simulation above
the mass threshold of $2.2\times10^{10}\hMsun$.  At $z=0$ there are
14,219 such halos.  The output for each halo includes the list of its
bound particles, the location and velocity of its center, and the NFW
profile parameters (e.g., $\cvir$ and $\Mvir$) and corresponding
errors ($\sigma_c$ and $\sigma_M$) when applicable. (We also include
information about the halo angular momentum properties; this is not
used in the present work, but see also \citealt{rw:01},
\citealt{vitvit:01} and Wechsler et al. in prep).  The mass function
of this revised halo catalog, and a detailed comparison with the halo
catalog used in the work of B01, has been presented in \citet{rw:01}.

\section{CONSTRUCTING A MERGER TREE}
\label{sec:tree}

As a base for the structural merger tree, we use the distinct halo
catalogs described above at each of 36 output times of the
simulations, from $z=7$ to the present.  The output times are well spaced
in redshift; the cosmological scale factors associated with the saved
output times are shown in Figure \ref{fig:time}.  Between each set of
output times, we track the trajectories of all of the
particles.  Given a halo and an output time, we tag all of its particles
and track them back to the previous output time.  We then make a list of
all halos at that earlier output time containing tagged particles,
recording the number of tagged particles contained in each one.  A
similar list of halos and tagged particle numbers is obtained for the
neighboring future output time.  In addition, we record the number of
particles that are not in any halo in the previous output time, and the
number of particles that do not end up in a halo in the subsequent
output time.

\begin{figure*}
\begin{center}
\resizebox{0.48\textwidth}{!}{\includegraphics{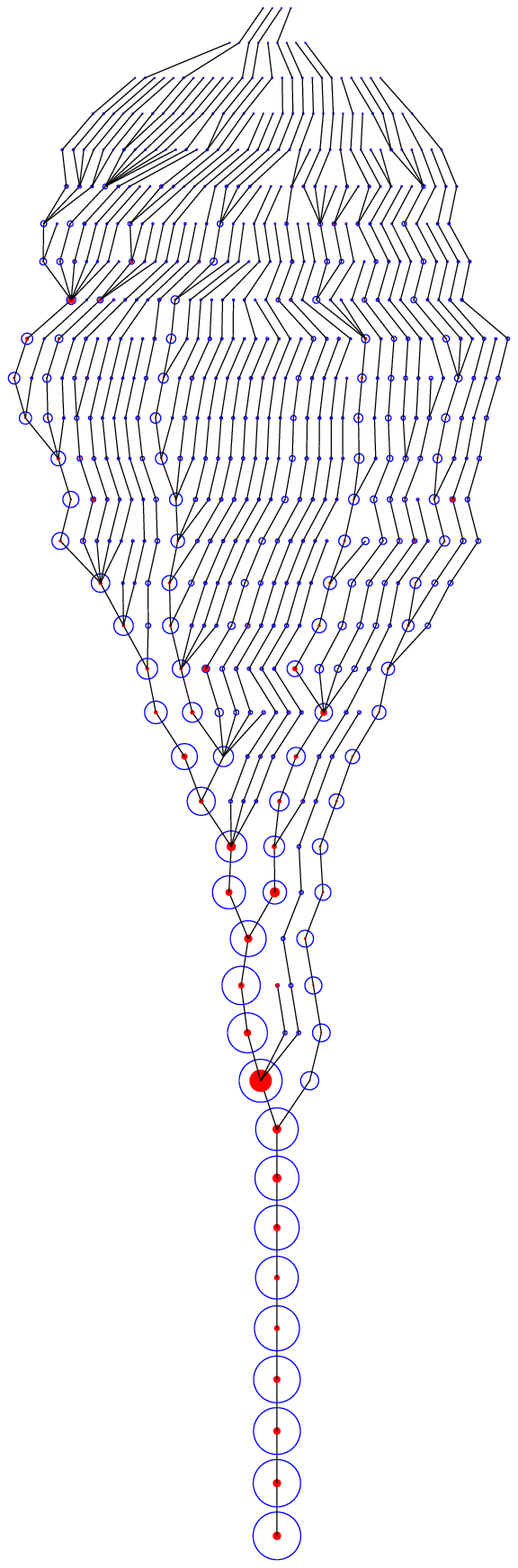}} 
\resizebox{0.48\textwidth}{!}{\includegraphics{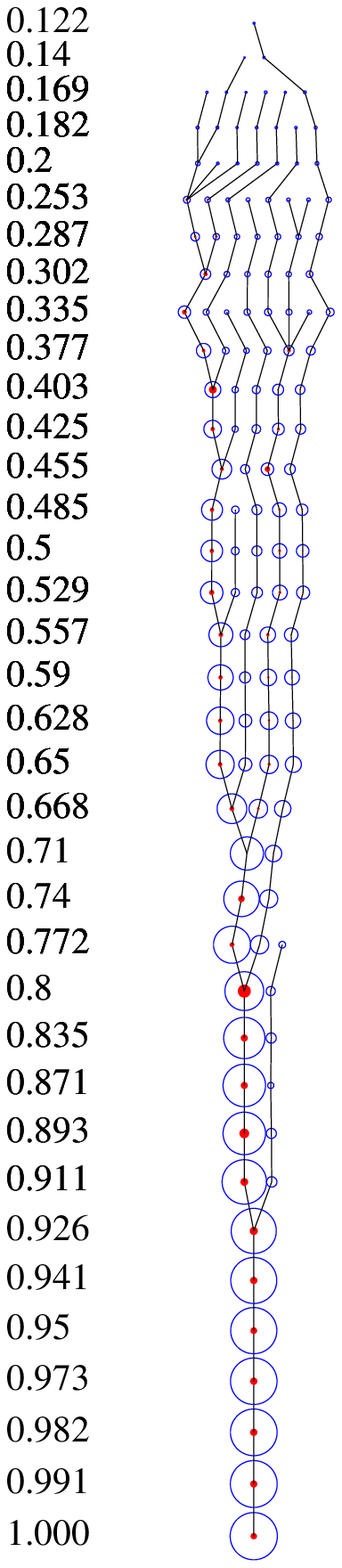}} 
\end{center}
\caption{
Structural merger trees for two halos. This diagram illustrates
the merging history of a cluster-mass halo (left;
$\Mvir=2.8 \times 10^{14} \hMsun$ and $\cv=5.9$) and a 
galaxy-mass halo (right; $\Mvir=2.9 \times 10^{12} \hMsun$ and
$\cv=12.5$) at $a=1$. The radii of the outer and inner (filled)
circles are proportional to the virial and inner NFW radii, $\rvir$
and \rs, respectively, scaled such that the two halos have equal sizes at
$a=1$. Lines connect halos with their progenitor halos. All
progenitors with profile fits ($M>2.2 \times 10^{11} \hMsun$) are
shown for the cluster-mass halo; all progenitors ($M>2.2 \times 10^{10}
\hMsun$) are shown for the galaxy-mass halo The scalefactor $a$ at the
output time is listed in the center of the plot.  The width of the
diagram is arbitrary.
\label{fig:tree}  
}
\end{figure*}

We have two criteria for halo 1 at one output time to be labeled a
``progenitor'' of halo 2 at the subsequent output time.  In our
language, halo 2 will then be labeled an ``offspring'' of halo 1.
First, more than half the particles in halo 1 must end up in halo 2.
In addition, since, on occasion, a halo's particles do not end up in
any halo in the subsequent output time, we also demand that more than
70\% of the particles in halo 1 which end up in any halo must end up
in halo 2.  Thus a halo is allowed to have only one offspring, but
there is no limit on the number of progenitors a halo may have.  In
much of the work that follows, we will focus on a ``most massive
progenitor'' halo in the previous output time.  In general, this is
the progenitor halo which contributes the largest number of particles;
however if the halo's progenitor mass is not at least half of its
mass, we additionally require that the progenitor's most bound
particle is included in the halo.  Even if this is the case, we also
check that the halo's most bound particle in the present output time
was a member of the progenitor; if it was not, the halo is only
designated as the most massive progenitor if it is more massive than
the minimum mass of the halo catalog and it is least a third the mass
of the offspring halo.  These criteria were designed to maximize the
redshift extent of as many mass accretion trajectories as possible,
without counting trajectories which may have been affected by the
completeness of the catalog or severe errors in fitting.

We have used the criteria outlined above to construct the merger tree
of every halo at every output time.  Examples of such a merger tree are
shown in Figure \ref{fig:tree}, which shows the mass accumulation
history of a small cluster halo of mass 
$\Mvir=2.8 \times 10^{14} \hMsun$ at $z=0$, and a galaxy-mass halo
($\Mvir=2.9 \times 10^{12} \hMsun$ at $z=0$).  Each halo in the tree
is represented by two circles, one with radius proportional to the
halo's virial radius and one with radius proportional to the halo's
best-fit NFW $\rs$ parameter.

\begin{figure*}
\centerline{\epsfxsize=0.9\textwidth\epsffile{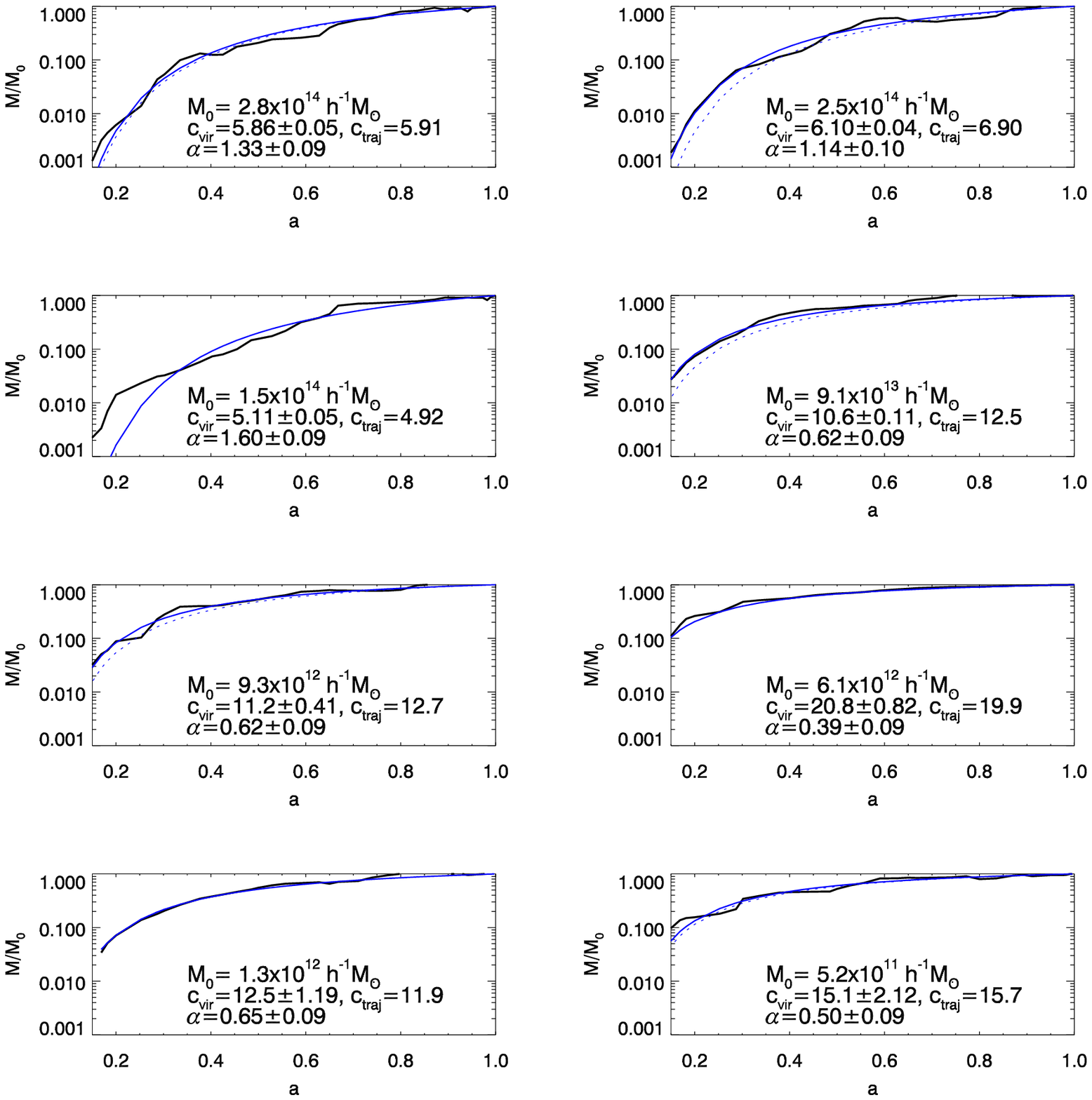}} 
\caption	
{Selected mass accretion trajectories, showing the evolution of the
most massive progenitor for individual halos in the simulation
(thick).
Functional fits to the growth curve of each halo using
Eq. \ref{eq:fit} are shown as thin smooth lines.  The dotted lines
represent the expected halo trajectory based on the value of the
concentration parameter, using Eq.~\ref{eq:fit}, with the value of $\ac$ as
derived by Eq.~\ref{eq:c} using the value $\cvir$ measured at $z=0$
(and quoted in each panel). The quoted $c_{\rm traj}$ is derived by
Eq.~\ref{eq:c} using the $\ac$ of the best fit to the actual growth
curve.
\label{fig:traj}}
\end{figure*}

\section{MASS GROWTH CURVES}
\label{sec:mah}

For the purposes of understanding the evolution of halo
concentrations, it is useful to characterize the history of mass
assembly in each halo by tracking the evolution of its most massive
progenitor.  The mass assigned to the most massive progenitor at each
output time is typically the best fit virial mass, $\Mvir$.  However,
for cases in which the fit errors on $\Mvir$ were large, we used an
iterative procedure, described in the appendix, to determine a
corrected mass; this mass is based on either the fit mass, the
measured mass within $\Rvir$, or an interpolated mass between the
previous and subsequent timesteps.  In addition, our halo mass
trajectories do not always extend as far back as the first analyzed
output time of the simulation. This usually happens because the most
massive progenitor at some redshift falls below our completeness limit
and cannot be identified, although there are also rare cases in which
our criteria for progenitor are simply not met (see \citealt{rw:01}).
In order to have complete trajectories for a reasonable number of
halos, and in order to have reliable fits for most halos, we limit our
analysis to halos more massive than $ 10^{12} \hmsun$ at $z=0$.  In
this mass range, which includes $\sim 900$ halos, $\sim 95\%$ of the
halo trajectories extend back to $z=2$, and $\sim 90\%$ extend back to
$z=3$.

\begin{figure*} 
\resizebox{0.47\textwidth}{!}{\includegraphics{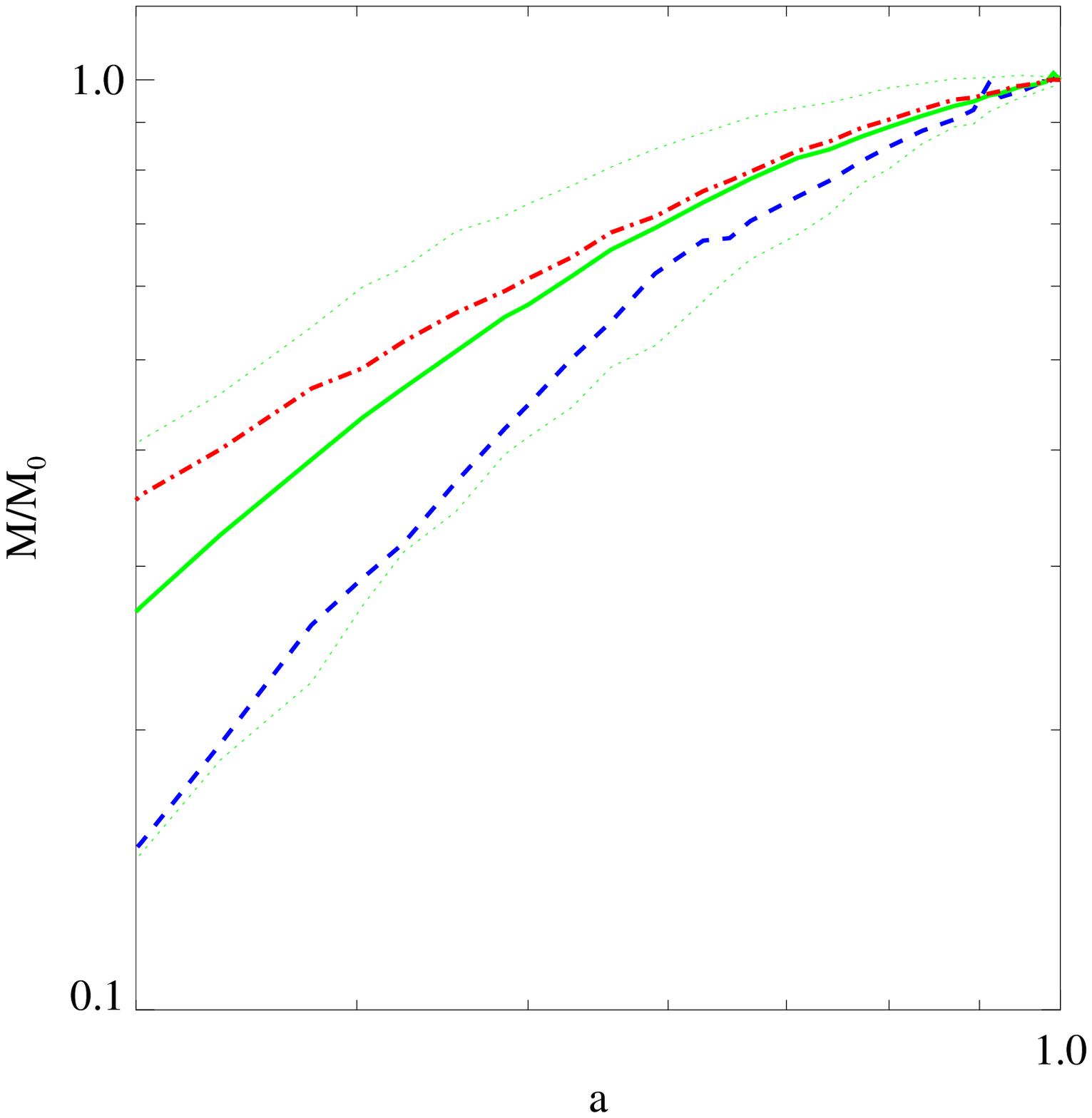}} 
\resizebox{0.47\textwidth}{!}{\includegraphics{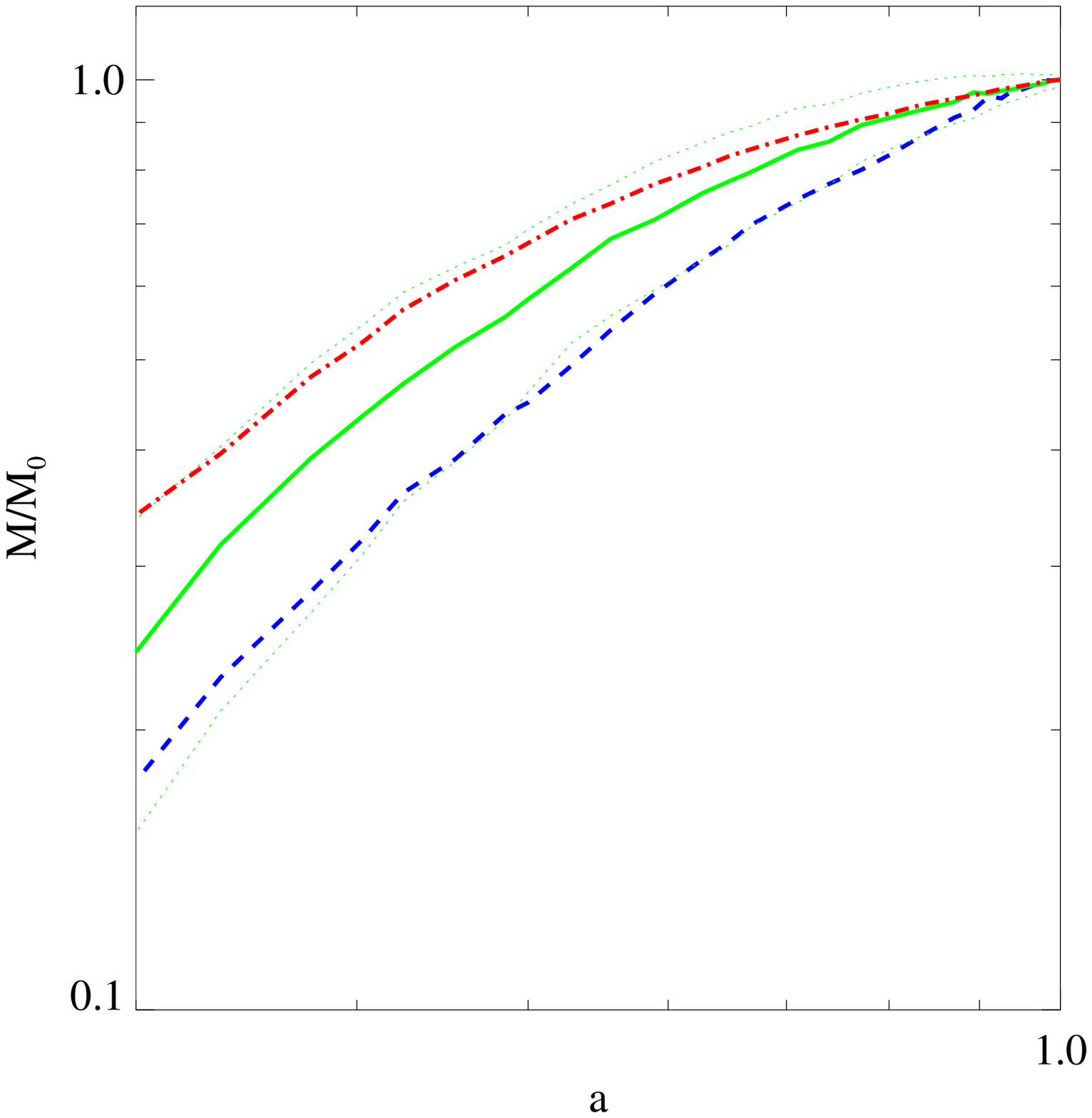}} 
\caption{
Average mass accretion histories, normalized at $a=1$. 
{\bf Left}: binned in 3 bins by final halo mass:
$M_0=1-4 \times 10^{12} \hmsun$ (dot-dashed),
$M_0=.4-3\times 10^{13} \hmsun$ (solid), and $M_0 > 3 \times 10^{13}
\hmsun$ (dashed).
The three (green) curves connect the averages of $M(a)/M_0$ at each
output time.  The pair of dotted lines shows the 68\% spread about the
middle case (the spread is comparable for the other bins).  We see
that massive halos tend to form later than lower mass halos, whose
mass accretion rate peaks at an earlier time.  {\bf Right}: binned in
3 bins by formation epoch $\ac$.
Dot-dashed lines correspond to early formers (typically low mass
halos), dashed lines to late formers (typically higher mass halos).
The averages and spread are displayed in analogy to the left panel.
\label{fig:avetraj1}
}
\end{figure*}

Figure~\ref{fig:traj} shows the history of mass growth for the major
progenitors of several different halos, spanning a range of masses
and concentration parameters.
Notice that more massive halos tend show substantial mass accumulation
up to late times, but the growth curves for less massive halos tend to
flatten out earlier.  This tendency is illustrated in Figure
\ref{fig:avetraj1}, which shows the average mass accretion histories
for halos binned by final mass.  Again, the high mass halos form later
than low mass halos, as expected in CDM models.

Since mass accretion is a continuous process, the loose term
``formation time'' is ambiguous and it requires an agreed, measurable,
quantitative definition.  The trends of formation time with mass and
redshift may depend on this definition.  One common choice is to set
the formation time equal to the time when the mass in progenitor halos
(or in the most massive progenitor) is equal to some fraction of the
halo's final mass $\Mobs$ \citep[e.g., NFW 1997,][]{lacey:93}.
Definitions of this type have a common feature: the formation time is
a relative measure which depends, for a given halo trajectory, on the
redshift $\zobs$ at which the halo is observed.  As $\zobs$ is
increased, the formation redshift $\zf$ also increases.  However, it
increases more slowly, since mass accretion proceeds more rapidly at
high redshift in CDM models.  Thus both the formation redshift $\zf$
and the ratio of $(1+\zf)/(1+\zobs)$ change with time in such a model.
As mentioned in the introduction, this feature of the formation time
definition is the reason that the redshift dependence proposed for the
NFW model fails to match accurately the dependence seen in
simulations.  Another shortcoming of this kind of definition is that
it is limited to using the value of the halo trajectory at one time,
which may introduce noise and miss relevant information.  It would be
useful to find a quantity which more fully characterizes the whole
assembly history of the halo, and preferably one which does not depend
on the redshift of observation $\zobs$ (at a possible cost that such a
definition may be allowed to have formation times in the future).

By examining a range of full mass assembly histories for our sample of
halos, we have found a useful parameterized form that captures many
essential aspects of halo growth over time.  Remarkably, we find that
both average mass accretion histories and mass accretion histories for
individual halos, as observed at $z=0$, can be characterized by a
simple function:
\beq
M(a) =  \Mobs e^{-\alpha z}, \quad a=(1+z)^{-1}.
\label{eq:fit}
\eeq
Although individual halo trajectories may deviate from this form
significantly in places (e.g., at the time of a major merger), this
one-parameter model (in addition to the halos' final mass $\Mobs$)
provides a remarkably good characterization of the range of halo mass
accretion trajectories.  Fits to this equation are shown in Figure
\ref{fig:traj} for several representative individual halos.
\citet{vdb:01} has independently shown that a similar, two-parameter, 
functional form can be used to represent halo mass accretion histories
for a variety of cosmologies and over a large mass range.

The single free parameter in the model, $\alpha$, can be related to a
characteristic epoch for formation, \ac, defined as the expansion
scale factor $a$ when the logarithmic slope of the accretion rate,
${\rm d}\log M/ {\rm d}\log a$,
falls below some specified value, $S$.  The functional form defined in
Eq. \ref{eq:fit} implies $\ac = \alpha/S$.

The same formation epoch can be defined equivalently for 
any ``observing'' epoch $\zobs$ of that halo,
by replacing $a$ in Eq.~\ref{eq:fit} by $a/\aobs$.
In this case, the characteristic formation time is 
related to $\alpha$ via
\beq
   \ac =\aobs\alpha/S.
\label{eq:ac}
\eeq
Thus at any such observing redshift, with scalefactor $\aobs=1/(1+\zobs)$ 
and mass $\Mobs=M(\zobs)$, the mass growth is fit by
\beq
M(a) = \Mobs {\rm exp} \left[-\ac S \left(\frac{\aobs}{a}-1\right)\right].
\label{eq:fit2}
\eeq
This implies that, for any halo whose mass accretion trajectory
resembles Eq. \ref{eq:fit}, the characteristic formation time is the
same regardless of the redshift $\zobs$ at which the halo is observed.
In what follows we have chosen $S=2$.  Since the value of $S$ is not
used for the fit but only serves to define \ac, this choice is arbitrary.
\footnote[4]{However, S=2 or something similar is required to match the B01 model
behavior that we describe in \S \ref{sec:formtime}.}

The range of formation scalefactors defined in this manner exhibits a
log-normal distribution, whose mean value increases with increasing
mass --- low mass halos typically accrete their mass early, while high
mass halos are typically still accreting mass rapidly in the present
epoch.  However, for a given mass, there is a large scatter in
formation epoch.  The scatter in the mass accretion trajectories for a
given mass can be seen in Figure \ref{fig:avetraj1}; we resume a
detailed discussion of the mass dependence of $\ac$ and the scatter
about this relation in \S \ref{sec:formtime}.

\begin{figure*} 
\resizebox{0.47\textwidth}{!}{\includegraphics{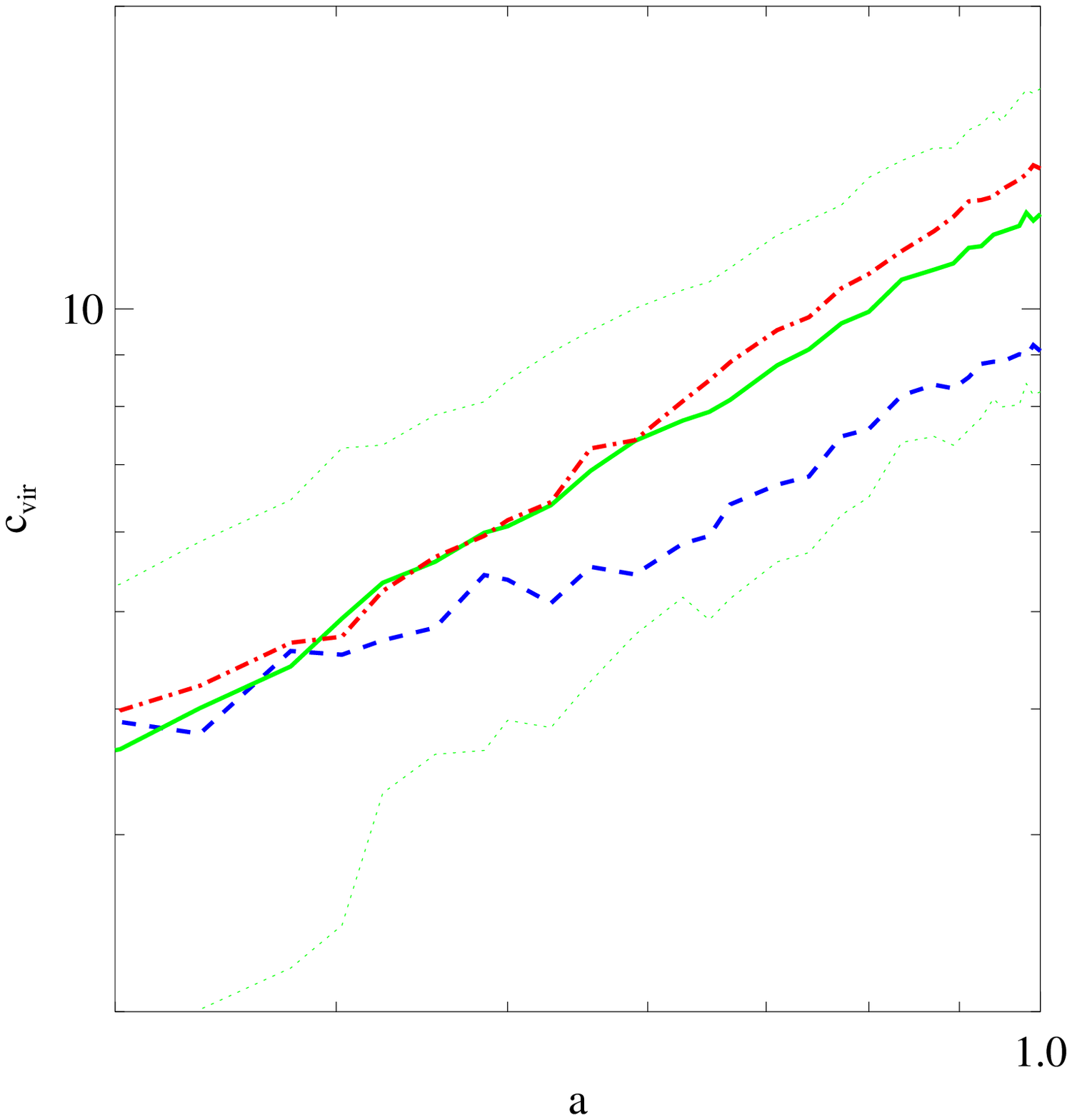}} 
\resizebox{0.47\textwidth}{!}{\includegraphics{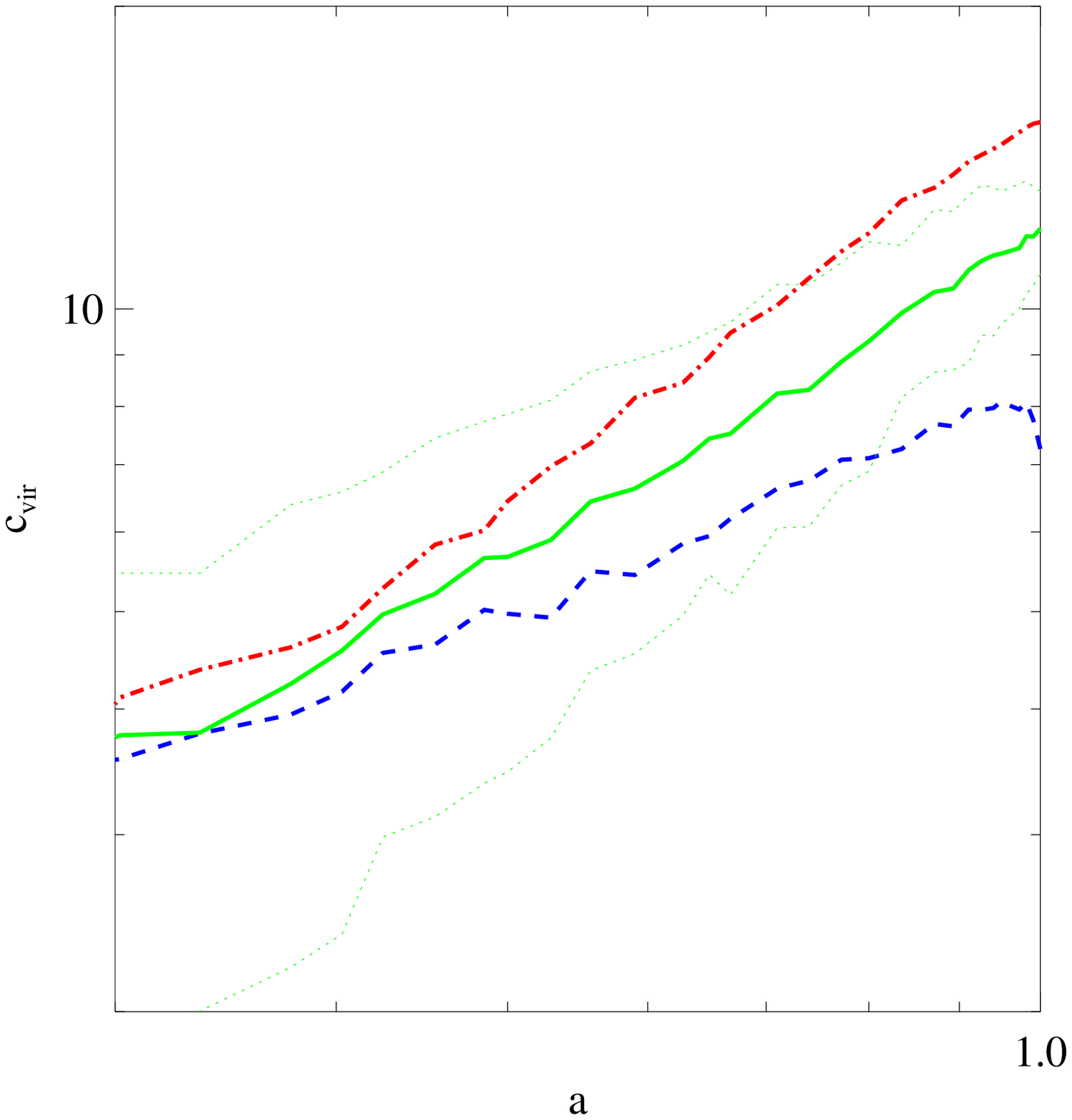}} 
\caption{Average evolution of the concentration parameter.
{\bf Left}: binned by the final halo mass as in the left panel of
Figure \ref{fig:avetraj1}. We see that massive halos typically have
lower, more slowly evolving \cv\ values than low mass halos.  {\bf
Right}: binned as a function of formation epoch as in the right panel
of Figure \ref{fig:avetraj1}.
\label{fig:avectraj}}
\end{figure*}

\section{CONCENTRATION AND ASSEMBLY HISTORY}
\label{sec:corr}
We find that the concentration of a halo is tightly correlated with
the characteristic formation epoch as defined in the above section.
Figure \ref{fig:avectraj} shows the average evolution of the
concentration parameter for halos in different mass ranges,
corresponding to the average mass trajectories shown in Figure
\ref{fig:avetraj1}.  From this figure it is clear that halo
concentrations have a stronger trend with less scatter when binned on
$\ac$ (right) than when binned on mass 
(left). \footnote{Note that the figure only shows this directly for
$z=0$, although it is true at any redshift $z_{\rm o}$ when $\ac$ is measured
at $z_{\rm o}$.  However, the scatter about the {\em average trajectory}
increases with $z$, since a halo can't uniquely predict its future.}
We therefore investigate how \cv\ is related to \ac\ directly.

\begin{figure*}[t]
\begin{minipage}[t]{0.48\linewidth}
\centerline{\epsfxsize=\colwidth\epsffile{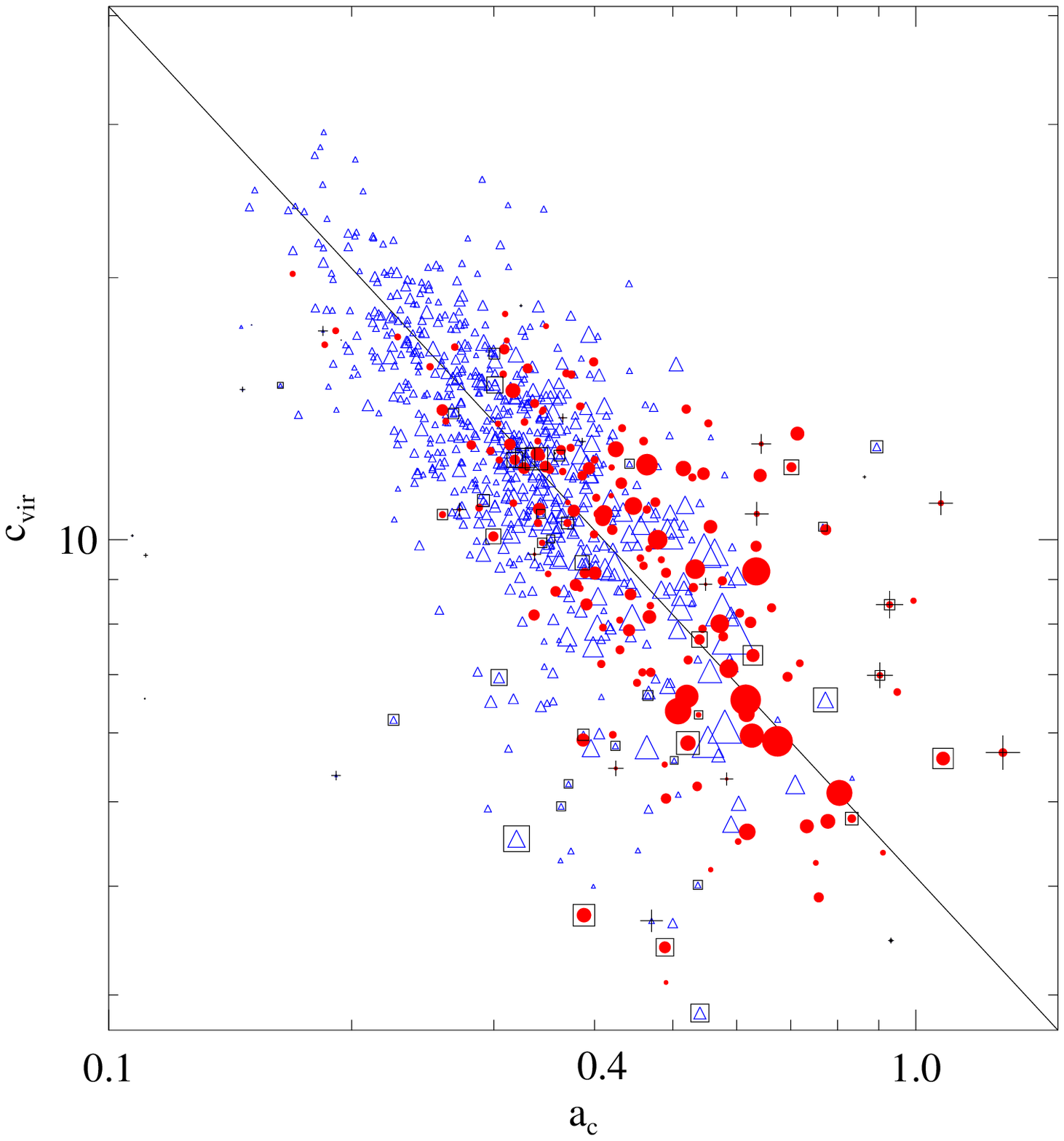}}
\caption
{Concentration versus formation epoch for halos more massive than
$10^{12} \hMsun$ at $z=0$. Halos that had major mergers ($M_2$ >
$M_1/3$) since $z=1$ are indicated with filled circles, halos that did
not have major mergers since $z=1$ are shown with open triangles.  The
size of the points is inversely proportional to the error on the
points (in both
\cv\ and \ac).  Halos with short trajectories are marked with pluses
and excluded from further analysis; halos whose final $c_vir$ value jumps are
outlines with squares and treated as described in the text.
\label{fig:cacz0}}
\end{minipage}
\hfill
\begin{minipage}[t]{0.48\linewidth}
\centerline{\epsfxsize=\colwidth\epsffile{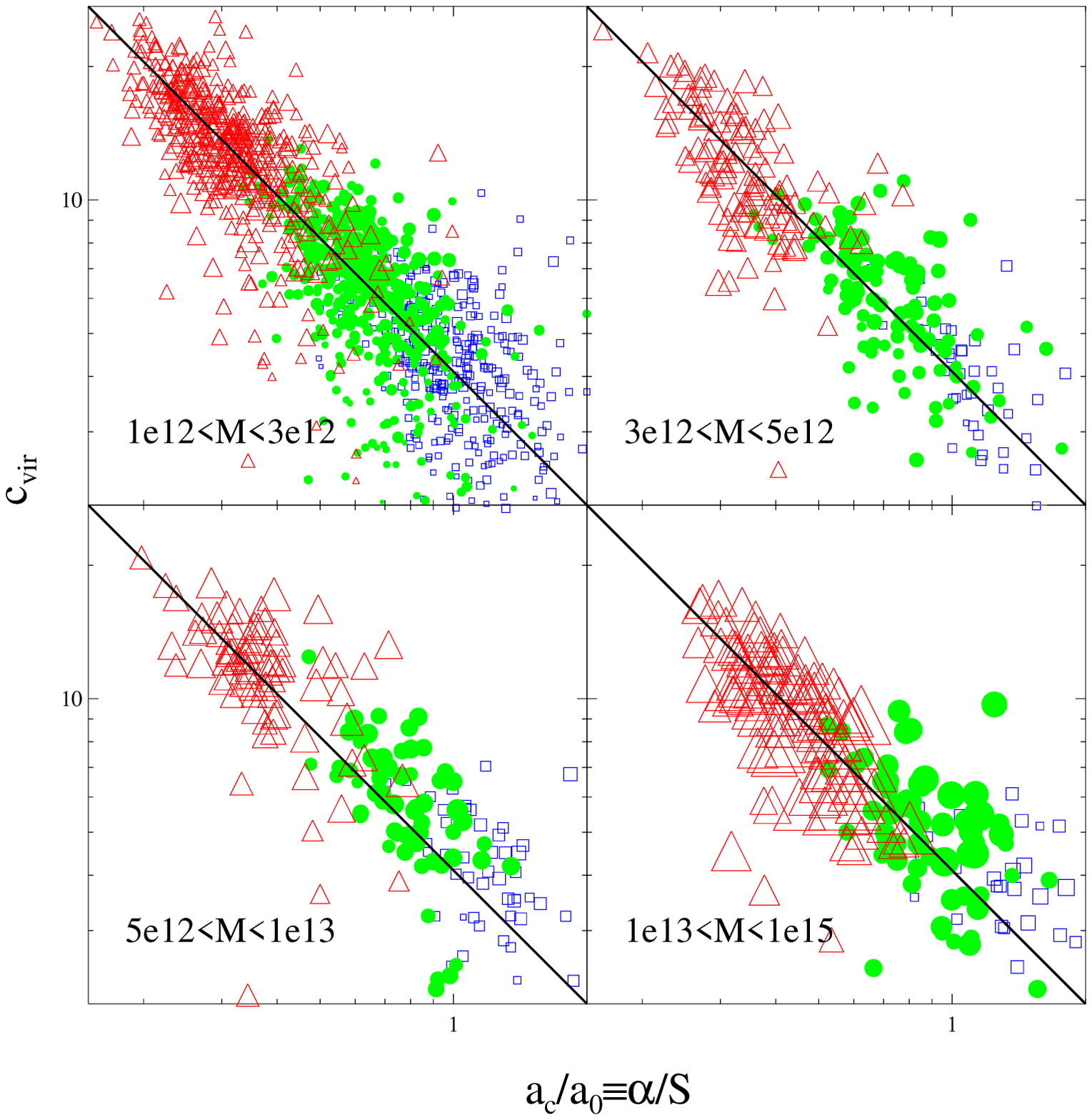}} 
\caption
{Concentration versus scaled formation epoch $\ac/\aobs$, for halos at
$z=0$ (triangles), $z=1$ (circles), and $z=2$ (squares).  The 4 panels
correspond to different mass ranges.  At all masses and redshifts, the
concentration parameter $\cv$ is well fit by the functional form $\cv
= \cc\ac/\aobs$, where $\cc \sim 4.1$ (represented by the solid line
in each panel).
\label{fig:cac}}
\end{minipage}
\end{figure*}

Figure \ref{fig:cacz0} shows the relation between concentration and
$\ac$ for halos at $z=0$.  The concentration of a halo is strongly
correlated with its characteristic formation time,
and a good fit is obtained with the inverse relation: 
\beq
\cvir = \cc/\ac , 
\eeq 
where $\cc=4.1$ is the typical concentration of halos forming today.  
The scatter about this relation is already not too
large for all the halos, but we note that most outliers fall in one of
the following three special categories:
\begin{enumerate}
\item 
the halo has a truncated trajectory
that does not extend far back to the past, 
and thus $\ac$ is not well determined;
\item 
the halo has a significant discontinuity in its \cv\ trajectory at the
final output time only, so that this value is not representative of the
whole trajectory (this can occur if there is a merger or other
disruption occurring at the final output time); or
\item 
the assembly history includes a merger that is substantially larger than
the average accretion rate at that time, and thus Eq. \ref{eq:fit}
does not provide a good description of the actual history.
\end{enumerate}
To deal with special case (1), halos with trajectories which do not
extend as far back as $z=1$ are excluded from further analysis (fewer
than $5\%$ of cases; these halos are indicated by plus symbols in
Figure \ref{fig:cacz0}).  For case (2), outlined by squares in Figure
\ref{fig:cacz0}, we find that a much better agreement with the median
relation is obtained when the last discrepant value of $\cvir$ is
replaced by the value of $\cvir$ in the preceding output time.  We do
not attempt to cure the problem associated with case (3), except for
keeping in mind that the outliers remaining in the $\cv$-$\ac$
relation are often due to a failure of Eq.~\ref{eq:fit} to adequately
model the history of that halo.  With these modifications, the scatter
in $\cv$ for a given $\ac$ is $\Delta({\rm log}\cv) \approx 0.09$,
without removing additional scatter due to large NFW fit errors (see
also Figure \ref{fig:hist}), and $\Delta({\rm log}\cv) \approx
0.05-06$ when errors in $\cv$ are corrected for.

We have also examined the dependence of $\cv$ on the {\em merger}
history of halos, which is correlated with but distinct from the mass
accretion history of the most massive progenitor discussed above.
Since halos that did not undergo a recent major merger are more likely
to have accreted most of their mass early, they typically have earlier
formation times and higher concentration values (see Figure
\ref{fig:typescat}).  However, we find that the parameter $\ac$ we
have defined based on the mass assembly history is more useful; for
halos with a {\em given} $\ac$ value, the occurrence of a recent
merger is not an important factor affecting the concentration.  This
can be seen in Figure \ref{fig:cacz0}, which demonstrates that halos
which have experienced recent major mergers (circles; defined here as
a merger ratio of greater than 1/3) form later on average but follow
the same trend in \cv\ vs. \ac\ (albeit with more scatter; see \S
\ref{sec:scatter}) as halos with no recent major mergers (triangles).

\section{DEPENDENCE ON REDSHIFT OF MEASUREMENT}

We now investigate how the correlation found in the previous section
between $\cvir$ and $\ac$ as measured at $z=0$ behaves as a function
of the redshift $\zobs$.  We found that 
\cv\ is inversely proportional
to \ac,\ or equivalently to the fit parameter $\alpha$ from
Eq. \ref{eq:fit}.  If the determining factor in a halo's concentration
is the shape of its mass growth curve then we expect that a similar
dependence would hold for all redshifts.  Eq. \ref{eq:ac} would then
imply that the concentration should be inversely proportional to
$\ac/\aobs$ for any $\aobs$ when the halo is observed.  In Figure
\ref{fig:cac}, the measured concentration values are plotted vs. this
scaled formation epoch for halos at $\zobs = 0, 1,$ and $2$.  Halos
appear to follow the same trend regardless of mass or redshift.

\label{sec:acm}
\begin{figure*}[t]
\centerline{\epsfxsize=\textwidth\epsffile{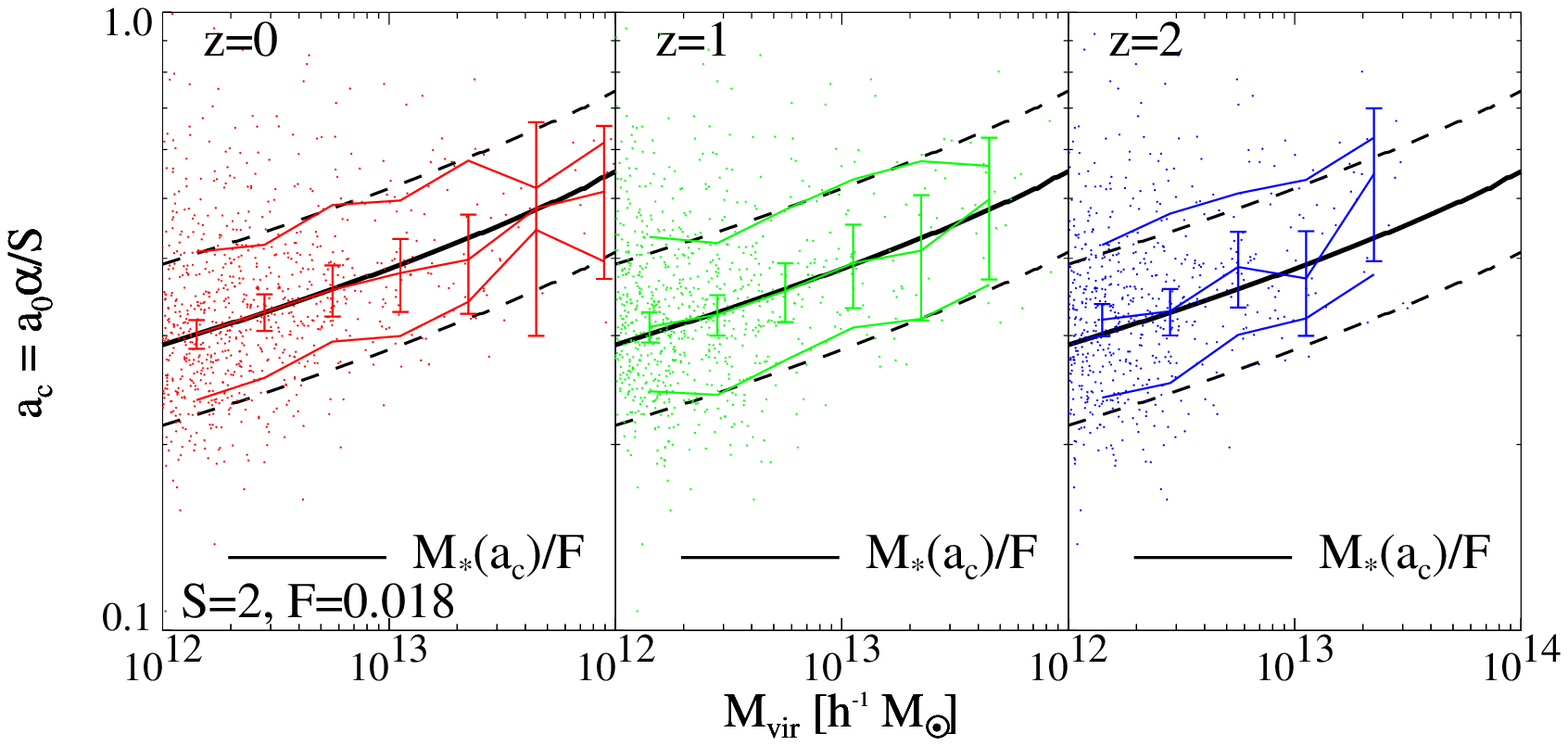}} 
\figcaption
{Mass dependence of $\ac$, shown for halos at redshifts $z=0$, $1$,
and $2$.  Dots represent individual halos.  The middle thin lines with
Poisson error bars indicate median values for the mass bins, and the
outer thin lines indicate the 68 \% scatter about the median.  In each
panel, the thick solid line is $\act$ as a function of mass from the
model of B01 and the thick dashed lines indicate the scatter about
this value needed to account for the 35\% measured scatter in
concentration.
\label{fig:acm}}
\end{figure*}

In order to obtain the most reliable estimate of the proportionality
constant $\cc$ in the linear regression of $\cvir$ and $\ac/\aobs$, we
use halos in the mass range $1-5\times10^{12} \hMsun$, and consider
the errors in both $\ac$ and $\cv$ while excluding outlying points
that are more than 2-$\sigma$ away from the best fit line.
Considering halos at $\zobs=0, 1$, and 2 together yields the same
result as obtained for $\zobs$ halos alone, properly scaled by
$\aobs$:
\beq
\cvir = \cc \aobs/\ac,  
\label{eq:c}
\eeq
where $\cc=8.2/S$ (here we have used $S=2\rightarrow\cc=4.1$.) 
As before, the
parameter $\cc$ is the typical concentration of halos whose formation
time is at the time of measurement, $a_c=\aobs$. For the $\Lambda$CDM
cosmology considered here, and $\aobs=1$, this is the typical
concentration of halos of $\sim 7\times10^{13} \hMsun$.  Figure
\ref{fig:cac} shows that this formula provides a good description
of the observed correlation between concentration and formation time
for halos at all masses and redshifts.

\section{PROPERTIES OF THE FORMATION TIME}
\label{sec:formtime}

As discussed above, B01 showed that the average concentration value
measured for halos of a fixed mass scales as $\cv \propto \aobs$.
This trend was understood using a simple model in which the central
densities of halos are set by the density of the universe at a
characteristic collapse time, $\act$, implying $\cvir \propto
\aobs/\act$.  Crucial to the success of this model was that the
definition of collapse time, $\act$, was independent of $a$ for fixed
mass halos.  In the previous sections we showed that a similar result
obtains using a different definition of the characteristic formation
time: $\cvir \propto \aobs/\ac$, where \ac\ is derived based on the
actual history of mass accretion in individual halos, rather than on a
simple universal scaling argument as in B01.  In order to obtain a
consistent trend with redshift, we expect that \ac\ (like $\act$) will
be independent of the epoch $\zobs$ when the measurement is performed
for a given halo mass.  This assumption is tested in Figure
\ref{fig:acm}, which shows the dependence of $\ac$ on mass, for halos
identified at three distinct redshifts in the simulation.  Within the
errors, it shows roughly the same mass trend regardless of redshift;
this is a key feature that our $\ac$ parameter has in common with the
collapse epoch $\act$ defined by B01.  In fact, these parameters can
be directly associated (for an appropriately chosen value of $S$).
For example, we find that $\act$ follows the same mass trend as our
derived $\ac$ (with $S=2$) if it is defined as in B01 ($\Ms(\ac)=FM$)
with $F=0.015$.

\begin{figure*}[t]
\begin{minipage}[b]{0.48\linewidth}
\centerline{\epsfxsize=\colwidth\epsffile{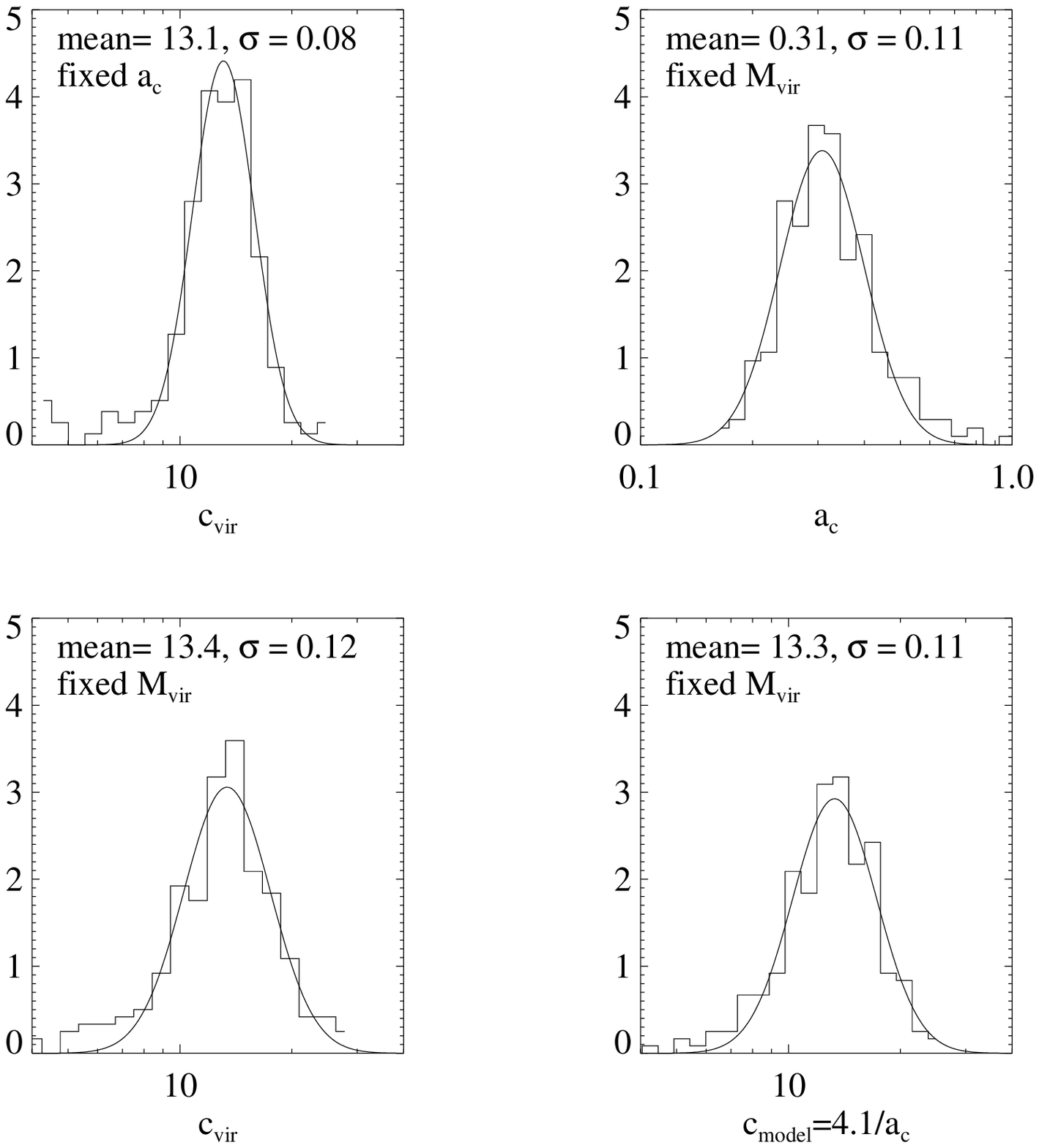}} 
\caption{
Probability distributions of \ac\ and \cv\ when other parameters are
fixed.  {\bf Top left}: the distribution of \cv\ for a given value of
\ac; the scatter is less than $\sim 20\%$ (with no correction for fit
errors in \cv\ or \ac), and less than $\sim 15\%$ when corrected for
errors in $\cv$. {\bf Top right:} the distribution of \ac\ for halos
in the mass range $1.5-2.5\times10^{12}\hmsun$. {\bf Bottom left}: the
distribution of \cv\ for halos in the same mass range.  {\bf Bottom
right}: the distribution of \cv\ for the same mass range, as predicted
from \ac\ via Eq. \ref{eq:c}.  The mean and scatter of each
distribution is listed at the top of the panel and is virtually
equivalent between the modeled \cv\ (bottom right) and the measured
simulation values (bottom left).
\label{fig:hist}
}
\end{minipage}
\hfill
\begin{minipage}[b]{0.48\linewidth}
\centerline{\epsfxsize=\colwidth\epsffile{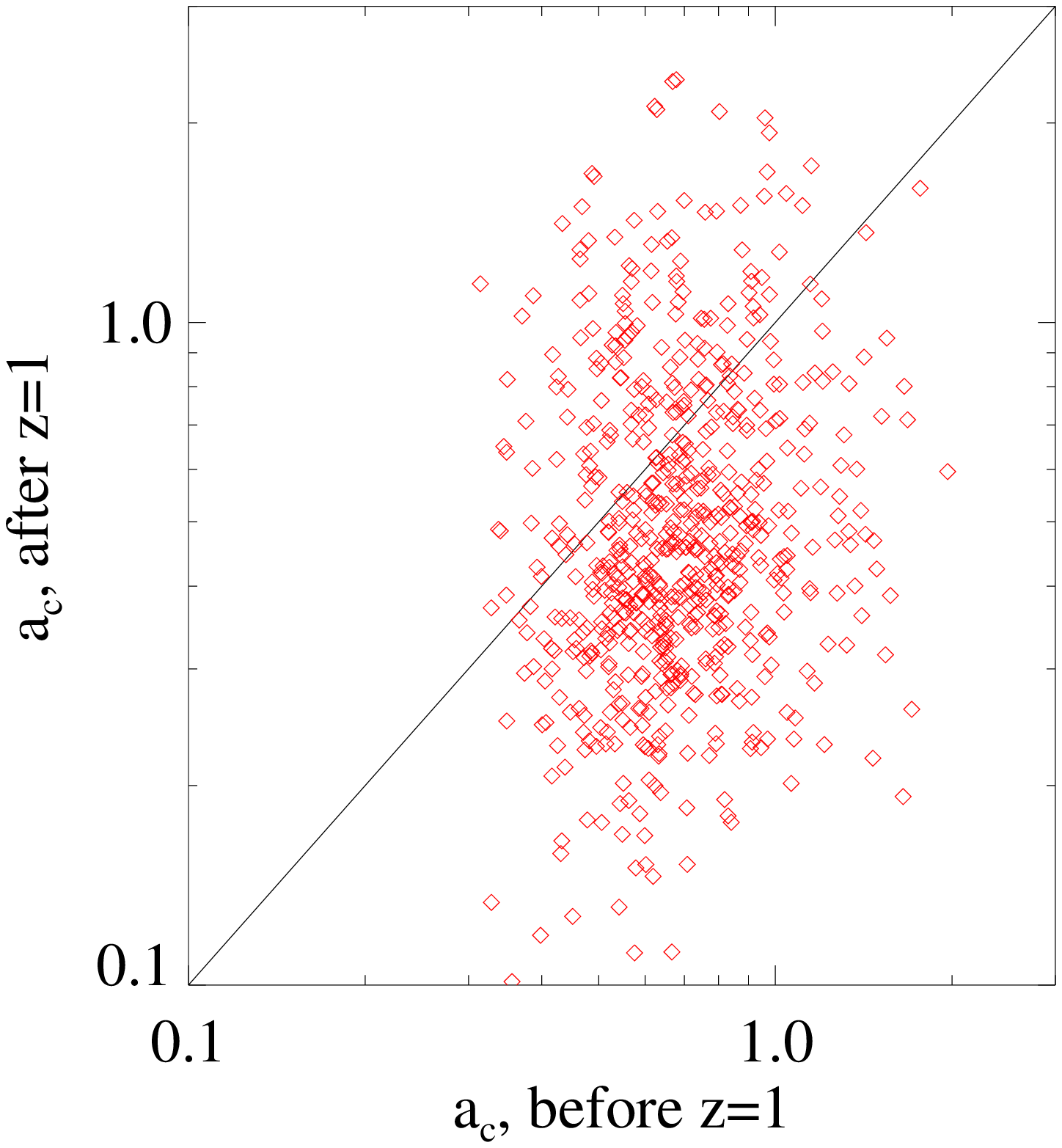}} 
\caption
{Comparison of the formation epoch $\ac$ as measured using
the trajectory to $z=1$ and the trajectory after $z=1$.
Either of these values individually is well correlated with the standard value
using the whole history (i.e., on average  $\ac$ stays constant for a given trajectory),
but they are not clearly correlated with each other,
indicating that a halo's past does not predict its future.
Note that the vertical scatter is larger than the horizontal scatter,
reflecting the fact that the second half of the trajectory does not constrain
the value of \ac\ very well because this is the shallow part of the function.
\label{fig:nocorr}
}
\end{minipage}
\end{figure*}

The value of \ac\ for a given mass halo also shows significant
scatter, as can be seen in Figure \ref{fig:acm}. The amount of scatter
($\sim 0.13$ in the log) can almost completely account for the scatter
seen in the \cv\ vs. $M$ relation (discussed in detail in \S
\ref{sec:scatter}).  Figure \ref{fig:hist} shows probability
distributions of $\ac$ and $\cv$ for a given mass range.  The
intrinsic scatter in $\cv$ for a given \ac\ is relatively small
compared to the scatter in $\cv$ for a given mass, and we find that
the measured distribution of $\cvir$ can be practically reproduced if
Eq. \ref{eq:c} is used to transform each measured $\ac$ value into a
concentration.  Thus the scatter in concentration can be understood as
deriving almost exclusively from the range of accretion histories for
a given halo mass.

It is encouraging that our definition of formation time is robust both
in terms of measuring the same value at different epochs along the
growth curve of a given halo, and in terms of its average value, for
halos of a given mass when measured at different redshifts.  We
explore this further in the next section.  Although a given halo mass
accretion history that is a perfect fit to Eq. \ref{eq:fit} will have
a constant value of \ac\ regardless of when it is observed, we find
that {\em the early part of a halo's trajectory is not a good
indicator of the latter part}.  This is shown in Figure
\ref{fig:nocorr}, which demonstrates that the formation time measured
using the first half of a halo's history is not correlated with the
formation time measured using the second half.

\section{REPRODUCING THE RESULTS WITH EPS} 
\label{sec:eps}

We have demonstrated that the tight correlation found in the simulations
between \ac\ and $\cvir$ can account for the mass and redshift trends
and the scatter in $\cv(M)$.  However, it would be useful to have a
way to model the concentrations without such a computationally
expensive simulation, for incorporation into analytic or semi-analytic
models.  Of course, the model of B01 did this to some extent for
{\em average} halo properties as a function of mass and redshift, but
this neglects the tight correlation found here between {\em
individual} halo density profiles and their mass accretion histories.
In this section we compare the mass accretion trajectories from our
simulation with trajectories generated using the EPS formalism and
test whether the correlation we found, $\cvir \propto \ac^{-1}$, can
be used to predict concentrations for {\em individual} halos
semi-analytically.

Comparisons of specific aspects of mass accretion histories as derived
semi-analytically and as extracted from simulations have been
performed (SLKD; \citealt{gardner:00}), but here we focus on the
quantity relevant for our modeling of $\cvir$, namely the range of
values of $\ac(M,z)$, which has not been investigated previously.
There are essentially four relevant questions in this analysis:

\begin{itemize}
\item do EPS trees produce the same range of \ac\ values for a given mass
as halo trees extracted from simulations?

\item do EPS trees produce the same trend of 
$\ac$ with $M$, 
which could then explain the 
trend of $\cv$ with $M$?
\item is $\ac(M)$ constant with redshift for EPS trees, which could then
explain the $\cv(z)$ trend?
\item does the scatter in $\ac$ for a given mass in EPS trees account
for the scatter in the measured $\cv(M)$ relation?
\end{itemize}

\subsection{$\ac$ vs $\mvir$ and the EPS Offset} 

In order to answer these questions, we first generate a random
ensemble of mass accretion histories.
This is done by coupling the extended Press-Schechter formalism, which
predicts the probability of accreting a given mass in a given time,
with a method for generating Monte-Carlo merger trees.  Here we use
the scheme introduced by SK99, with an additional modification proposed
by \citet*{bkw:00}.  For completeness, we outline the fundamental
aspects of the method in Appendix \ref{sec:epsapp}.

We consider the list of all halos found in the simulation with masses
larger than the minimum-fit mass $2\times10^{11} \hmsun$ ($\sim 3000$ halos).
We then generate ten Monte-Carlo realizations of mass accretion
trajectories for each of these halos, based solely on its $z=0$ mass,
keeping track of the growth of the most massive progenitor as a
function of time.  We start at $z=0$ and trace histories back to
$z=7.2$.  Using this most massive progenitor trajectory, $\alpha$ is
calculated by fitting the trajectory to Eq. \ref{eq:fit}.  The value
of $\ac$ is then defined by Eq. \ref{eq:ac}, with $S=2$.

Figure \ref{fig:achist} shows the distribution of \ac\ values found
for different given mass ranges, for both the simulated halo
trajectories and the EPS trajectories.  We see that the distribution
of \ac\ values from EPS trees is systematically offset toward later
formation times for all mass ranges, and also appear to be slightly
broader.  A similar discrepancy was found by SLKD, although they did
not investigate the same quantity --- they found the average mass of
the largest progenitor to be larger in the EPS trees than in the
simulated trees at low redshift, and smaller at high redshift.  This
implies later formation times for the simulated halos, as seen here.
This discrepancy can be understood in terms of comparing the
conditional progenitor mass functions; SLKD showed that EPS
over-predicts it for low masses and under-predicts it for high masses,
with the mass scale of the crossover decreasing with increasing
redshift.  This discrepancy in formation epochs seems to directly
reflect the more well-known finding that PS over-predicts the number
density of halos below \Ms\ and under-predicts the number density
above \Ms\ \citep[e.g.,][]{gross:98, st:99}.  Although only one
cosmological model (\LCDM) has been investigated here, the
disagreement of the halo mass function with Press-Schechter is generic
and we expect that the formation time discrepancy would be seen to
some extent in any CDM cosmogony.

\begin{figure*}[t]
\begin{minipage}[b]{0.48\linewidth}

\centerline{\epsfxsize=\colwidth\epsffile{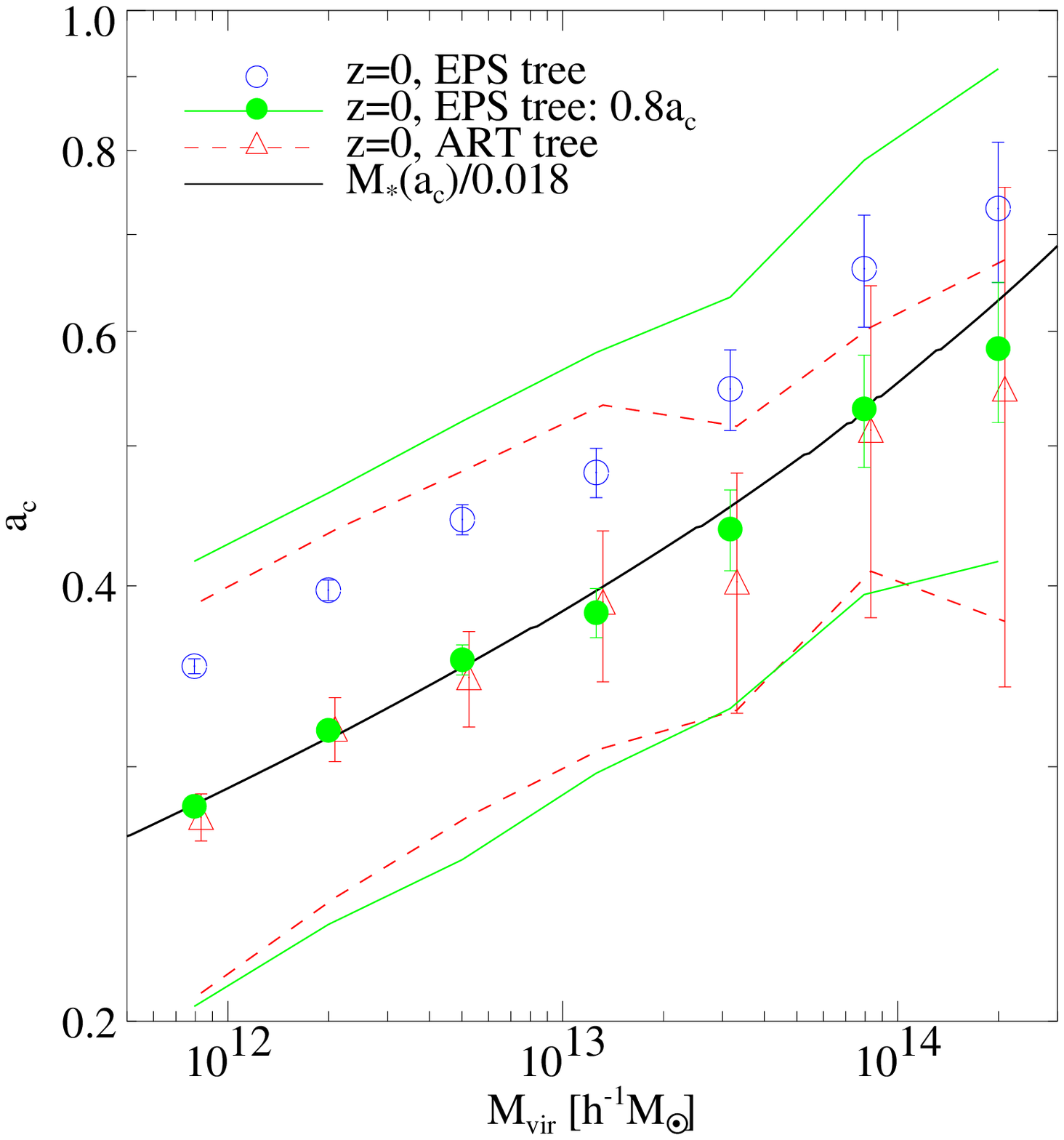}} 
\caption
{Comparison of the median $\ac(M)$ relation with that derived from EPS
trees.  The triangles correspond to the trees from the ART simulation.
The open circles represent ten Monte-Carlo realizations of EPS merger
trees for each simulated halo mass.  The EPS-derived $\ac$ values are
offset about 25\% higher than those measured in the simulation.  The
filled circles are the EPS results shifted down by 20\%.  In each case
the error bars are Poisson based on the number in the mass bin. The
outer lines represent the 68\% scatter about the median relation.  The
dark solid line in the middle represents the model of B01 with
F=0.015, and 35\% scatter about that model.
\label{fig:acmeps}
}

\end{minipage}
\hfill
\begin{minipage}[b]{0.48\linewidth}
\centerline{\epsfxsize=\colwidth\epsffile{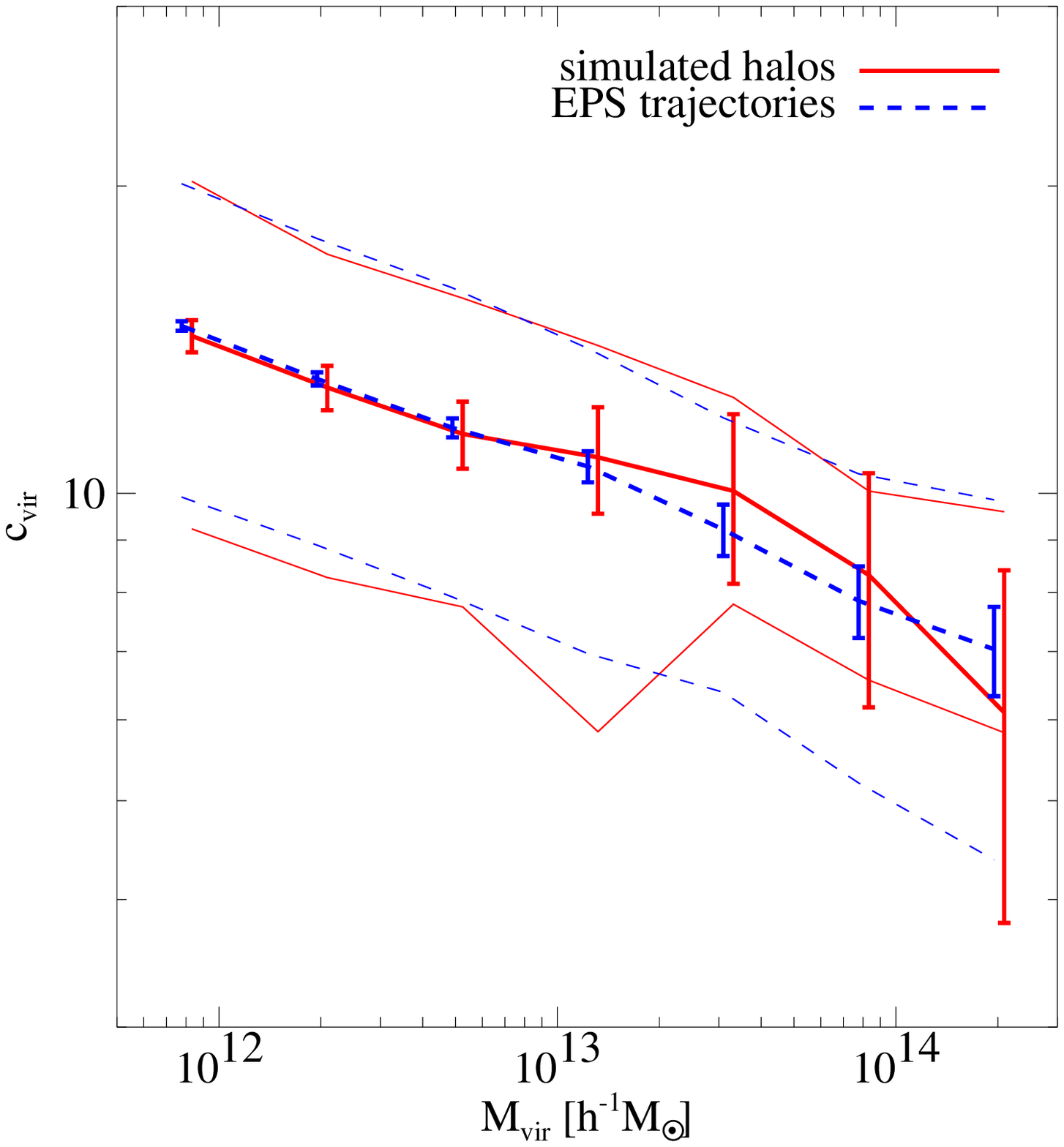}} 
\caption
{Comparison of the median $\cvir(M)$ relation 
for simulated halos (solid lines) with those obtained
using Eq. \ref{eq:c} and the \ac\ values derived from EPS trajectories
(dashed lines).
As in  Fig. \ref{fig:acmeps}, the EPS \ac\ values have been offset 20\% lower.
Ten Monte-Carlo merger tree realizations are generated for each
simulated halo mass.  Some selected realizations look virtually
identical to the distribution for simulated halos.  
The outer pair of thin lines
represent the corresponding 68\% scatter and  error bars are Poisson based 
on the number of halos in each bin.
\label{fig:cm_eps}
}
\end{minipage}
\end{figure*}

Although we find a discrepancy in the median \ac\ values as a function
of mass, we find that they are offset by a constant multiplicative
factor.  Figure \ref{fig:acmeps} shows the median and 68\% scatter for
$\ac(M)$, from the EPS trajectories and the simulated trajectories,
for $z=0$ halos.  The constant offset is in the sense that the
characteristic formation epochs derived from the EPS trajectories are
roughly 25\% larger than those measured from the simulation trees.
The scatter is quite comparable, though it is slightly larger for 

\begin{inlinefigure}
\centerline{\epsfxsize=\colwidth\epsffile{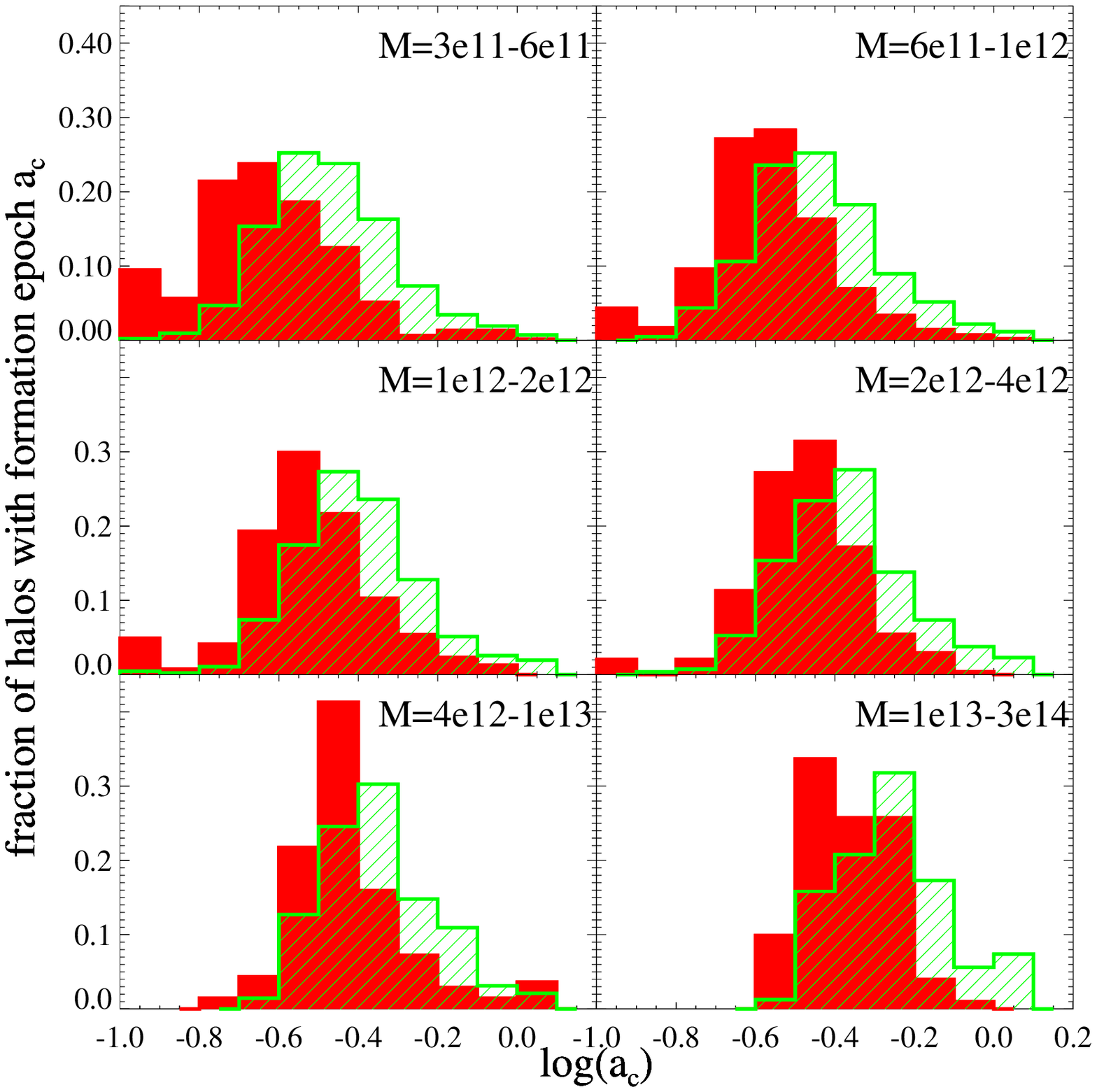}} 
\figcaption
{Comparison of $\ac$ values in the simulation with those derived from
EPS trees, for various mass ranges in the different panels.  
The filled histogram represents
simulated halos and the shaded histogram represents ten 
realizations of EPS trajectories for the same set of masses.  
Abnormal halos (whose trajectories end prematurely) in the simulations
are counted in the left-most bin; the number of
these cases is small, and negligible above $10^{12} \hmsun$.  
\label{fig:achist}
}
\end{inlinefigure}

\noindent the EPS-derived values.

If we assume that the \ac\ values found using EPS trajectories are
correct except for this constant 25\% offset, due to the known
discrepancies in PS theory, then these values can be used in
combination with Eq. \ref{eq:c} to estimate \cvir\ for each halo,
using its EPS-derived trajectory.  Figure \ref{fig:cm_eps} shows the
measured \cvir(\mvir) for simulated halos at $z=0$, compared with the
\cvir\ value obtained using this semi-analytic method.  The constant $\cc$
has been shifted by a factor of 1.25, but otherwise we are able to
match both the mass trend and the scatter of the measured \cvir\
values in the simulation with this simple model.

\subsection{Redshift Dependence in EPS}

\begin{figure*}[t]
\begin{minipage}[b]{0.48\linewidth}
\centerline{\epsfxsize=\colwidth\epsffile{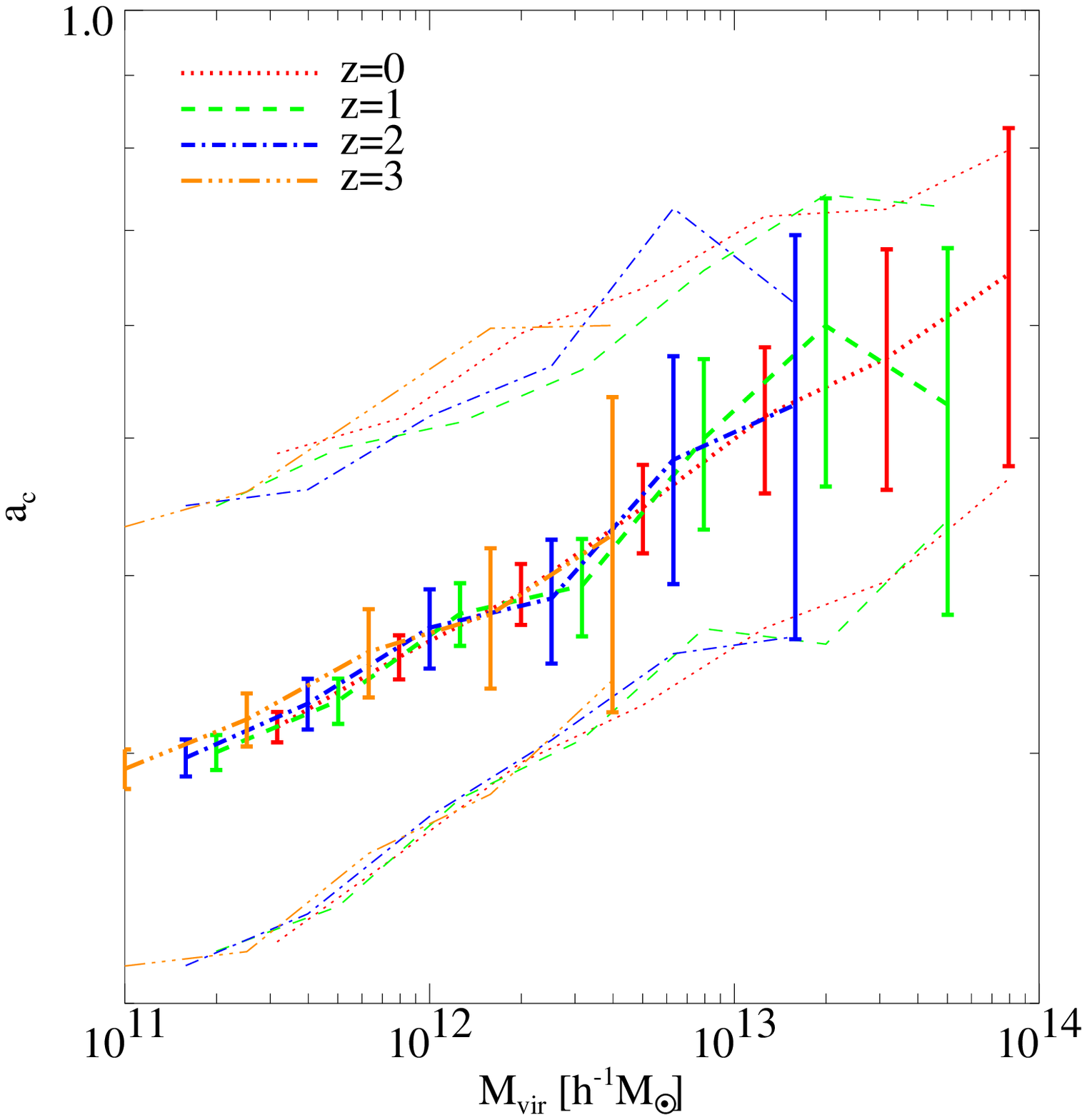}} 
\caption
{Robustness of $\ac(M)$ to the redshift of ``observation''
in one realization of an EPS tree.  
The 4 different sets of curves and symbols refer to 4 different redshifts.
The thick lines represent the median \ac\ values and the pairs of outer thin
lines represent 68\% scatter.  Error bars represent Poisson error based on the number in each
mass bin.  
As predicted by Eq. \ref{eq:fit}, the median
\ac\ value is uniquely determined by the halo mass, independently of
the redshift.  
\label{fig:acz_eps}
}
\end{minipage}
\hfill
\begin{minipage}[b]{0.48\linewidth}
\centerline{\epsfxsize=\colwidth\epsffile{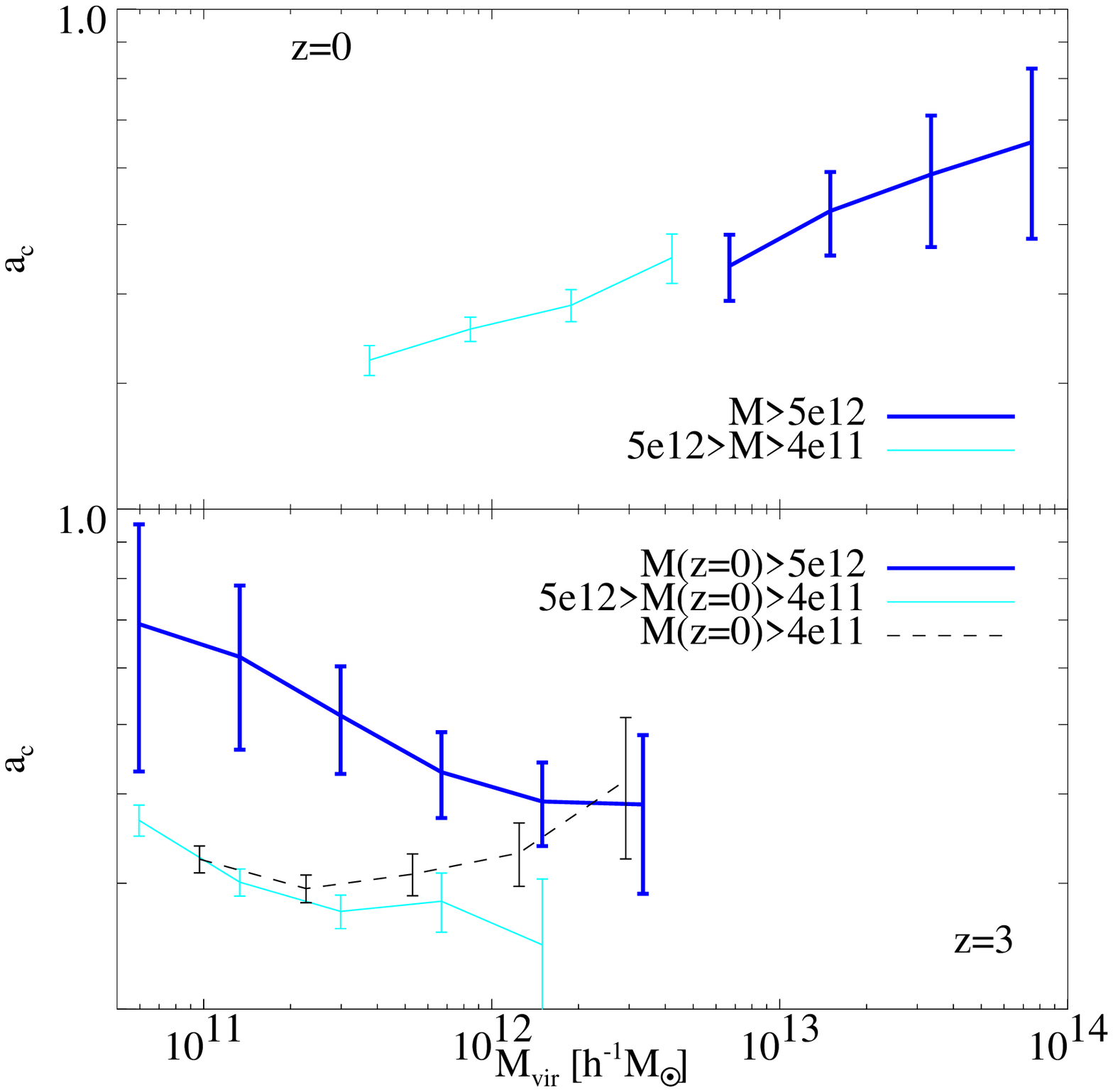}} 
\caption
{
An analysis of the robustness of $\ac(M)$ to redshift, in EPS trajectories.
{\bf Top}:
$\ac(M)$ at $z=0$, broken into two mass ranges.  
{\bf Bottom}: $\ac(M)$ at $z=3$ for the same mass ranges determined at $z=0$ 
(dark thick: high mass; light thin: low mass).  
The ratio  M(z=3)/M(z=0) is related to $\ac$, such that for a given $z=0$ 
mass, halos with high  $\ac$ will have lower $z=3$ masses.
The dashed line shows the sum of the mass bins, which follows a similar trend
with mass as the $z=0$ halos.  
The apparent upturn in the left-most bin of the dashed curve
is due to the exclusion of halos less massive than $4\times10^{11} \hMsun$
at $z=0$, which contribute lower \ac\ values.
\label{fig:zdep}
}
\end{minipage}
\end{figure*}

We can also test whether the redshift dependence of the EPS
trajectories is consistent with our model, and with the \cvir\ trend
seen in the simulation.  B01 showed that for a given mass, $\cv \propto a$,
implying, for our model, that (on average) \ac\ must be uniquely set by the
mass, independent of redshift.  Figure \ref{fig:acz_eps} shows
the $\ac(M)$ relation at redshifts $z=0,1,2,$ and $3$, using one
realization of EPS trajectories (again, with the mass weighting of the
simulated halos).  There is no discernible change in the median or
scatter over the entire mass range examined.  Using Eq. \ref{eq:c} to
translate these values into predicted \cvir\ values will thus result
in exactly the trend found by B01, i.e., that $\cv \propto a$, with the
correct mass trend and scatter for all redshifts.

Given the fact that the \ac\ values are robust to changes in $\zobs$,
it might be considered somewhat puzzling that the $\ac(M)$ relation
remains constant with redshift, since clearly the halo masses are
significantly lower at high redshift.  Naively, one might think that
since the masses shift, the average halo of a given mass would have a
higher $\ac$ value in the past.  However, for a given mass halo at
$z=0$, those halos which have the latest formation times will have the
smallest masses at high redshift, and those which form earliest will
have the highest mass at high redshift.  When halos are combined with
the proper mass weighting, they thus conspire to keep the same
relation regardless of redshift.  Figure \ref{fig:zdep} demonstrates
this in detail.  In the upper panel, the $\ac(M)$ trend is shown at
$z=0$ for halos in two distinct mass ranges.  In the lower panel,
these same halos are shown at $z=3$: the $\ac$ values for each halo
remain essentially unchanged, but they are now plotted against their
$z=3$ masses.  Halos in any given mass range at $z=3$ consist of a
combination number of low \ac\ halos that will have low $z=0$ masses
and high \ac\ halos that will have high $z=0$ masses.  These add in a
manner that maintains the shape and normalization of $\ac(M)$.
The fact that $\ac(M)$ is constant with redshift is an indication that
the model proposed by B01 is a reasonable one --- i.e., that (on
average) the formation time of a halo is set only by its mass, and can
be related to the time that mass was a fixed fraction of \Ms.

In summary, the formation times of halos in EPS (using an improved
version of the SK99 method to generate merger trees) are somewhat
later than those found in the ART simulation.  However, if we measure
\ac\ values for these semi-analytic trajectories (using
Eq. \ref{eq:fit2}), multiply these values by a constant factor (0.8),
and translate these formation epochs to concentrations using the
relation $\cvir = \cc
\aobs/\ac$, we are able, to a good approximation, to match the
scatter, mass, and redshift dependence of the \cv\ values found for
simulated halos.  This method can be used in semi-analytic models to
estimate halo concentrations that are both based on their individual
mass growth histories and have the correct distribution at every
redshift.

The only disadvantage of the above method is that it requires
generating full mass accretion histories for a sample of halos.  Since
Eq. \ref{eq:fit2} is a one-parameter model, one might be tempted to
directly calculate the more prevalent formation redshift $z_{\rm{f}
0.5}$, when the most massive progenitor mass was half of $M_0$, and
whose probability distribution can be calculated directly from EPS
without generating merger trees, and then translate this value into
\zc\ using Eq. \ref{eq:fit2} in order to derive a concentration.
However, while this method predicts the mean values relatively well,
in fact there is substantially more scatter in values of $z_{\rm{f}
0.5}$ than in values of \zc.  This is due to the fact that individual
trajectories are noisy, and at any one point in the trajectory, there
is likely to be significantly more scatter than in the shape of the
halo as a whole.

\section{SCATTER IN THE \cvir(\mvir) RELATION}
\label{sec:scatter}

\subsection{Evaluating a Corrected Scatter}

The scatter in the concentration parameter has been estimated by B01
and \citet{jing:00}, and it may have a number of important
observational implications.  
\citet{jing:00} found a scatter of
$0.08 - 0.1$ in $\log_{10}\cvir$, while B01 derived a somewhat larger
scatter of $\Delta \log_{10}\cvir = 0.14$.~\footnote[5]{The value of
$0.18$ reported in the abstract of B01 was actually the directly
measured scatter, without any correction for fit errors.  The
corrected value of $0.14$ is illustrated in B01 Figure 4.  This
footnote is aimed at correcting this oversight in that paper.}  
A revised scatter estimate from our improved halo catalogs, which also
relies on the mass trajectories, is presented here.  We also discuss
how the scatter in $\cv$ is affected by the merging history of halos.

\begin{figure*}[t]
\begin{minipage}[b]{0.48\linewidth}
\centerline{\epsfxsize=\colwidth\epsffile{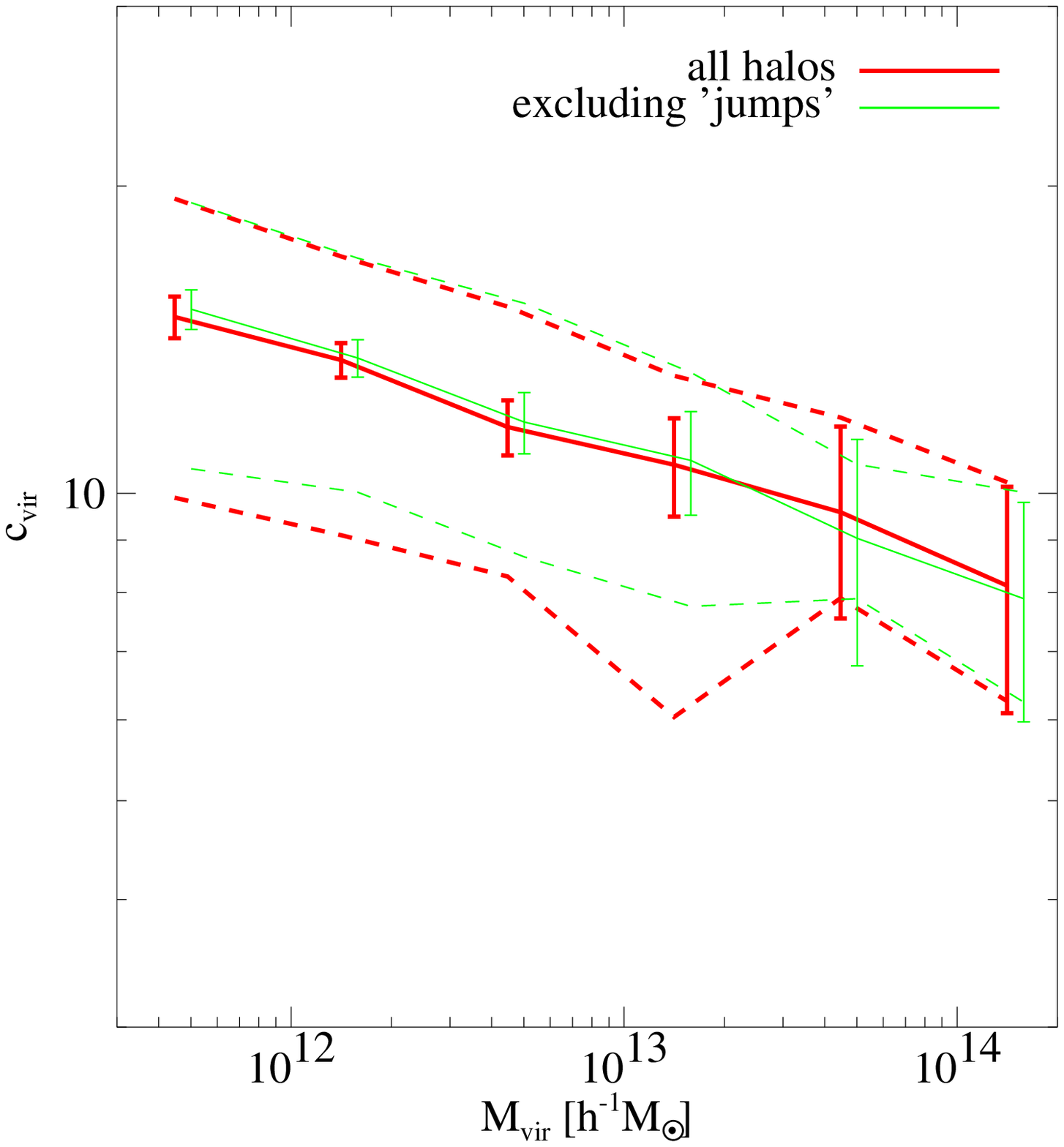}} 
\caption
{Scatter in the $\cvir$-$\mvir$ relation for halos at $z=0$.  The thick
lines represent all the halos.  The thin lines represent all halos 
whose concentration has not jumped by more
than a factor of two (in either direction) since the previous output time
(at $z=0.01$).  In each case, the solid lines represent the median
value of \cv, plotted with Poisson error bars, and the dashed
lines represent the scatter corrected for errors in the individual
profile fits and for Poisson scatter in the bins.  In the mass range
where the results are most reliable ($\sim 1\times10^{12} \hmsun$),
the value of the corrected scatter in these two samples is
$\Delta({\rm log}\cv)
\approx 0.14$ and 0.12, respectively.
\label{fig:cmscat}}
\end{minipage}
\hfill
\begin{minipage}[b]{0.48\linewidth}
\centerline{\epsfxsize=\colwidth\epsffile{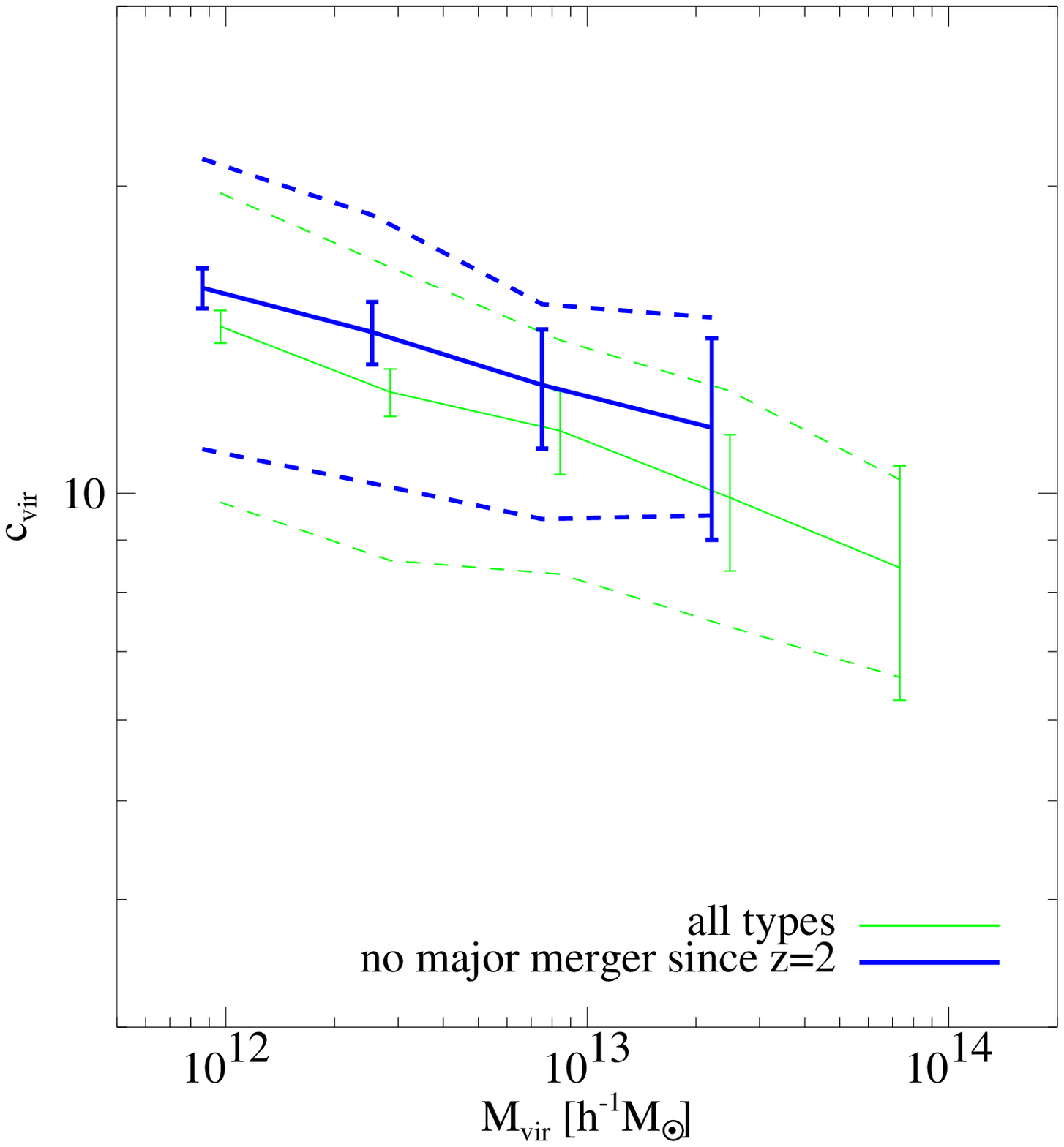}} 
\caption
{Type dependence of the $\cvir$-$\mvir$ relation for halos at $z=0$.
The thick lines represent halos which have not had a major merger
since $z=2$; the thin lines represent all types.  In each case halos
with big `jumps' in $\cv$ are excluded, as described in the previous
caption.  In each case, solid lines represent the median value of \cv,
plotted with Poisson error bars, and dashed lines represent the
intrinsic corrected scatter.  For halos without recent mergers, the
scatter is reduced to $\Delta({\rm log}\cv) \approx 0.10$.
\label{fig:typescat}}
\end{minipage}
\end{figure*}

B01 devised a method for correcting the scatter estimate for errors in
the fits, and actually plotted this corrected scatter in Figure 4.
Our new halo catalogs have significantly fewer halos with large fit
errors, and thus a smaller uncorrected scatter, but encouragingly,
when we use the method of B01 for correcting this for fit errors and
Poisson errors, we get almost identical results (shown in Figure
\ref{fig:cmscat}).  This method involves doing 500 Monte Carlo
realizations of each halo, in which their $\cv$ value is chosen from a
one-sided Gaussian deviate with a standard deviation error on the
measured $\cv$ value.  This value is then added or subtracted to the
measured value depending on whether the measured value was below or
above the median value for that mass.  Poisson errors due to finite
statistics are then subtracted in quadrature to obtain an estimate for
the intrinsic scatter --- which we also find to be $\Delta({\rm
log}\cv) \approx 0.14$.

As mentioned previously, there may be halos which for a short period
of time are not well-fit by an NFW density profile; in many cases this
occurs when a halo is undergoing merging or disruption.  In most
cases, this results in an artificially low value of \cv, which usually
persists only for one timestep.  In order to remove the effects of
badly estimated concentration parameters which persist only for a
short time, we use the merging history of a halo to identify those
cases where the concentration has jumped significantly since the
previous output time.  The \cvir$(M)$ relation excluding these halos
(which make up $\sim8\%$ of the total) is plotted in Figure
\ref{fig:cmscat}, compared to the whole sample; the scatter for this
sample is reduced from $\sim 38\%$ to $\sim 31\%$.  Note that the
exclusion of these halos does not change the median value
significantly.  We regard the scatter estimate from excluding these
halos as a lower limit.  With this correction, our analysis is closer
to the scatter estimate presented by \citet{jing:00}.  It is possible
that some of the remaining discrepancy could be due to an
underestimate of fit errors; the total scatter for the very massive
halos in our analysis is slightly smaller.  However, it should also be
pointed out that \citet{jing:00} only considered relaxed halos, and
our analysis contains halos with a full range of properties, including
those that have been recently disrupted --- and these add
significantly to the scatter.

\subsection{Scatter and Halo Merger History}

As discussed in \S \ref{sec:corr}, halos with recent major mergers display
the same trend between concentration and \ac\ (Eq. \ref{eq:fit}), but
with somewhat more scatter.  We show in Figure \ref{fig:typescat} that
these halos also follow the same basic trend with mass.  This figure
compares this trend and scatter in $\cvir(M)$ for all halos with that
seen in halos that have not had a major merger since $z=2$.  The
scatter is reduced from about 31(38)\% for all halos to $\sim
26(28)\%$ for those halos without recent major mergers, where the
first listed estimate is when halos with large $\cv$ jumps are
excluded, and the number in parenthesis includes all halos.  (The
implied correction for halos without recent mergers is not large since
most of the halos with jumps are in the process of merging.)  The
sample of halos which {\em have} had a major merger since $z=2$ do not
have reduced scatter compared to the whole sample.  Since halos with
recent major mergers have later formation times on average, when they
are excluded from the sample, the remaining halos have higher $\cv$
values by a factor of about $\sim 10\%$ compared to the complete
sample; this can also be seen in Figure
\ref{fig:typescat}. 

The amount of scatter in the concentration parameter for a given mass
is of particular interest because of its possible implications for
scatter in the Tully-Fisher relation (see, e.g., B01;
\citealt{vdb:00}), and there is particular interest in the amount of
scatter in the concentrations of spiral galaxies.  In many scenarios
for galaxy formation, major mergers destroy disks, and thus halos with
recent major mergers are unlikely to host spiral galaxies.  As
described above, not counting these halos reduces the scatter of the
whole sample and thus may reduce consequential scatter in Tully-Fisher
to a level that can be matched with observations.  Whether these
remaining halos host spiral galaxies may be influenced by how much
mass they have accreted since their last disruption; this could bring
the concentrations of halos hosting spiral galaxies lower than the
median shown here.  It should also be noted that in this analysis we
have only considered distinct halos.  To get a full estimate of
scatter and normalization for galactic halos we would have to include
subhalos in the analysis, which B01 have shown have somewhat larger
scatter in \cv.  We defer a full analysis of these issues to a later
work.

\section{DISCUSSION AND CONCLUSIONS}
\label{sec:c_conclude}

Making use of a large sample of dark halos simulated in a cosmological
volume, we have studied the relation between mass accretion history
and the density concentration of halos.  Remarkably, halo mass growth
curves (normalized to the final halo mass) can be accurately described
by a one parameter function, in which mass accretion occurs rapidly at
early times, and slows at late times.  The characteristic ``formation''
time, $a_c$, defined as the time when the log-mass infall rate drops
below a fixed value, fully defines the trajectory.  We find that the
value of $a_c$ for each halo trajectory is independent of the epoch at
which the halo is observed, $\aobs$.  In addition, the average value
of $a_c$ for halos of fixed mass is independent of redshift.

The NFW halo concentration parameter, $\cvir$, which, in combination
with the halo mass, uniquely sets the shape of the density profile, is
tightly related to $\ac$ via $\cvir = \cc \aobs /\ac$.  The central
density of a halo at fixed radius grows rapidly when the mass
accretion rate is high, and approaches a constant value as the mass
accretion slows.  This result is consistent with a picture in which
the mass accretion rate determines how far accreted mass makes it into
the center of a halo: for high mass accretion rates, accreted material
makes it far into the center of the halo, but as it slows new material
builds up on the outside.  Thus central densities of halos asymptote
to a value which is proportional to the density of the universe at the
time when the mass accretion rate slows. In late forming halos this
process is delayed and thus the final central density is lower.  To
demonstrate the model, in Figure \ref{fig:model} we plot the
evolution of several variables for an early and late forming halo: the
mass, concentration, log-slope of the mass accretion rate, scale radius
\rs, and the density within a fixed radius.  Each parameter is
calculated analytically; this is done using Eq.~\ref{eq:nfw}, which
specifies the density profile, Eq.~\ref{eq:fit2}, which specifies the
mass accretion history, and Eq.~\ref{eq:c}, which specifies the
relationship between them.

We showed that scatter in \cv\ for a given mass can be explained
almost exclusively by scatter in \ac\ for halos of that mass.  Thus
this model, based on the $\cvir$-$\ac$ correlation, captures
successfully the main properties of the concentration parameter,
including its mass dependence, redshift trend, and the scatter about
these relations.

The formation times derived from semi-analytic realizations of EPS
mass accretion histories were compared with those measured in the
N-body simulation, and found to be systematically larger by $25\%$.
This offset reflects known inaccuracies in the Press-Schechter
approximation.  By adjusting the formation times for this offset, we
have successfully used the $\cvir$-$\ac$ correlation to reproduce the
mass and redshift dependence of \cv, and the scatter about these
relations, using mass accretion histories derived from EPS. This
technique is likely to be very useful for inclusion in semi-analytic
models of galaxy formation, which so far have at best drawn \cv\
randomly from an assumed global probability distribution, with no
dependence on the halo's history.

We have presented an estimate for the scatter in \cv, with preliminary
hints for how this scatter may depend on galaxy type.  For halos of a
given mass that have not had a major merger, we estimate that the
scatter is as low as $\Delta({\rm log}\cv) \approx 0.1$.  This scatter
is roughly consistent with the scatter observed in the Tully-Fisher
relation.  However, we have not included subhalos in this estimate, so
our results only apply to isolated field galaxies; including subhalos
may increase the scatter.  It is also interesting that such halos
(those without recent major mergers, which could conceivably host disk
galaxies) are also somewhat more concentrated, which is perhaps
contrary to na\"{\i}ve expectations.

It is worth pointing out that this is not the only attempt to
connect the full mass accretion history of halos to their density
structure.  For example, \cite*{av:98}, building on the work of
\cite{zh:93}, used an analytic approach based on shell-by-shell mass
accretion to connect halo structure with mass accretion (see also
\citealt{rg:87}).  Although their 

\begin{inlinefigure}
\centerline{\epsfxsize=\textwidth\epsffile{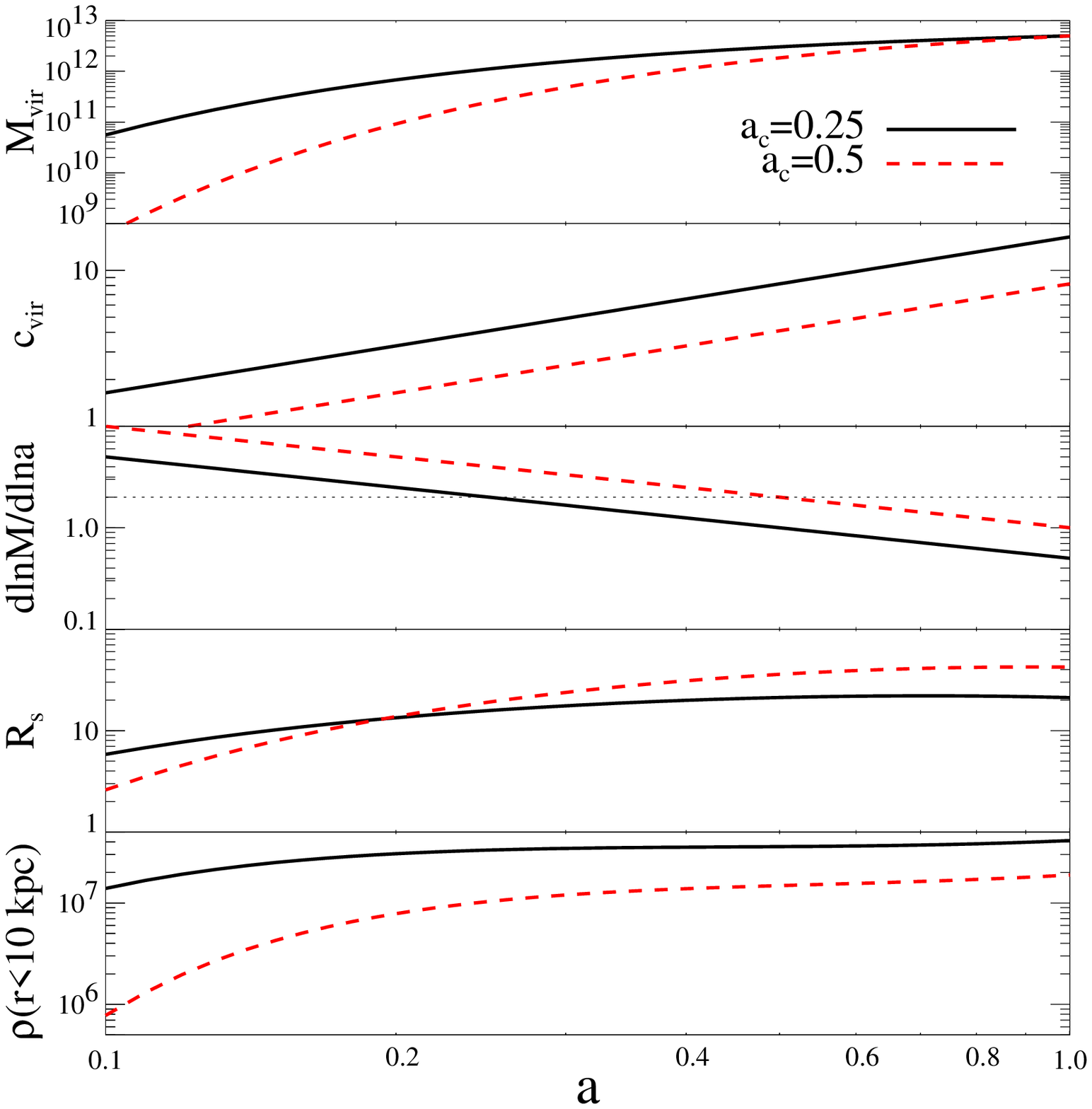}} 
\figcaption{
Evolution of mass and structure of two halos.  From top to bottom, we
plot the evolution of the mass $\Mvir$, the concentration \cv, the
log-rate of the mass accretion history dlogM/dloga, the scale radius
\rs, and the density of the halo within 10 kpc.  For each parameter the
evolution of two halos with the same final mass ($5 \times 10^{12}$)
but different values of \ac\ (0.25 and 0.5) is shown.  In the plot of
the mass accretion rate, the value of S used to define \ac\ is marked
with a thin dotted line.
\label{fig:model}  
}
\end{inlinefigure}

\noindent results differ in detail 
from what we have found using N-body simulations, the general trends
associated with early and late formation seem to agree.  Our results
may serve as a useful benchmark against which specific analytic and
semi-analytic models of halo structure formation can be tested.

The correlation we have found between individual halo assembly
histories and their density profiles is likely to have important
consequences for a large range of observable galaxy properties.  For
example, halo density profiles directly affect galaxy rotation curves,
and are likely to play an important role in determining galaxy shapes,
gas infall and star formation rates.  We thus expect that the
inclusion of such a correlation in the context of semi-analytic models
which track galaxy formation and evolution with simple recipes will
affect a number of predictions of these models, and thus will provide
a significantly more realistic theoretical framework for understanding
galaxy populations, the origin of galaxy type and the variation in
galaxy properties.

\section*{Acknowledgments} 
We thank Anatoly Klypin for useful discussions and for access to the
simulations used for this analysis, which were performed at NRL and
NCSA. We thank Tom Abel, George Blumenthal, Sandy Faber, Ari Maller,
Piero Madau, Rachel Somerville, Frank van den Bosch, and David
Weinberg for helpful discussions and comments.  In addition, we thank
the anonymous referee for several comments that improved the clarity
of the manuscript.  RHW received support from a GAANN fellowship at
UCSC.  JSB received support from NASA LTSA grant NAG5-3525 and NSF
grant AST-9802568.  AVK was supported by NASA through Hubble
Fellowship grant from the Space Telescope Science Institute, which is
operated by the Association of Universities for Research in Astronomy,
Inc., under NASA contract NAS5-26555.  This work was also supported by
NASA ATP grant NAG5-8218, NSF grant PHY-0070940 and a Faculty
Research Grant at UCSC, and by the US-Israel Bi-national Science
Foundation grant 98-00217.

\appendix
\section{Finding and Fitting Halos}

The details of our procedure for finding and fitting halos are as follows:

\begin{enumerate}
\item 
\label{item:one}
\ni
We  construct density field  values  by a Cloud-in-Cell (CIC)  process
\citep{hockney:81} on the  largest grid of  the  simulation $\Delta L$,
and rank the particles according to their local density as determined
on this grid.  We then search for the possible halo centers, using two
sets of smoothing spheres; one, with a small radius, $\rspo$, in order
to locate the centers of tight, small clumps; and the other, with a
larger radius, $\rspt$, in order to locate the centers of halos with
diffuse cores.  The larger radius, $\rspt$, is set equal to
$\Rvirmin$, the virial radius of a halo of mass $\Mvirmin$.  The
smaller radius is set to $\rspo = 5 \Rvirmin/\Npmin$, a rough
approximation to the radius within which our smallest halo would be
expected to contain $\sim 5$ particles.

For each set  of spheres, we  take from  the ranked list  the particle
with the highest local density and place  a sphere about its location.
A  second sphere is  placed  about the next  particle  in the list not
contained in the first sphere.  The  process is continued until all of
particles are  contained within at  least one sphere.  Because  we are
only interested in centers of halos  more massive than $\Mvirmin$, we
discard each sphere with fewer  than a set number  of particles.   The
minimum number  of particles required for  a kept sphere is determined
separately for each radius.

For  the  $\rspo$  spheres, we  use   the following  conservative halo
density profile:
\begin{eqnarray} 
  \rho(r) = \left\{\begin{array}{ll} 
                C/r^{2.5}_{\rm sp1} &\hspace{0.7cm}\mbox{$r<r_{\rm sp1}$} \\ 
                C/r^{2.5}  &\hspace{0.7cm}\mbox{$r>r_{\rm sp1}$},
                \end{array} 
        \right.
\end{eqnarray}
(where  $C$  is  determined by  fixing   the minimum halo mass   to be
$\Mvirmin$), in order to estimate the minimum number of particles
within $\rspo$:
\begin{eqnarray}
 \Nspo = \frac{\Npmin }
       {1  + 6[(\Rvirmin/\rsp)^{1/2} - 1]}.
\end{eqnarray}
For the $z=0$ output of the $60 h^{-1}$Mpc simulation we analyze,
$\Nspo = 3$ (rounding to the next lowest integer).  Spheres of size
$\rspo$ with fewer than $\Nspo$ particles are discarded.  Similarly,
all of the $\rspt$ spheres containing fewer than $\Nspt = \Npmin$
particles are discarded.

The final list of candidate   halo centers is  made up  of all of  the
(small)  $\rspo$ spheres, together  with  each of the $\rspt$  spheres
that \textit{does not} contain an $\rspo$ sphere.

\item  For each sphere of radius $r_{sp} = \rspo$ or $\rspt$, whichever
applies, we use the  particle distribution within the sphere to find  the center of mass
and iterate  until convergence.    We  repeat the procedure   using  a
smaller  radius,  $r=r_i$,  where $r_i=r_{sp}/2^{i   \over   2}$.   We
continue this method until $r_i = r_L$, where  $r_L$ is defined by the
criterion $r_{L} > 2 f_{\rm res} > r_{L+1}$, or until reduction leads to
a sphere with fewer than $\Nspo$ particles.

\item We unify the spheres whose centers are
within $r_{L}$ of each other.   The unification is performed by making
a  density weighted   guess for a  common center   of  mass, and  then
iterating to find a center of mass for  the unified object by counting
particles.  The size of sphere used to determine the center of mass is
the smallest radius that will allow the new sphere to entirely contain
both candidate halo spheres.
\item For each candidate halo center we step out in radial shells of 
$3 f_{res}$, counting enclosed particles, in order to find the outer
radius of the halo: $R_h \,=$ min($\rvir,\,R_t$).  The radius $\rvir$
is the virial radius, and $R_t$ is a ``truncation'' radius, defined as
the radius $(< \rvir)$ in which a rise in (spherical) density is
detected ($d \log\rho/d \log r > 0$).  This is our method for
estimating when a different halo starts to overlap with the current
halo and is important for halos in crowded regions.  We estimate the
significance of a measured upturn using the Poisson noise associated
with the number of particles in the radial bins considered.  Only if
the signal to noise of the upturn is larger than $\sigma_{R_t}$ do we
define a truncation radius. 
The value of $\sigma_{R_t}$ is a free parameter.  We use
$\sigma_{R_t} = 5$.~\footnote[6]{ The choice was motivated by several
tests using mock catalogs of halos in clusters designed to determine
how varying $\sigma_{R_t}$ affects our ability to fit the density
profiles of subhalos.  Although our results were not strongly
dependent on this choice, we did obtain the best fits using
$\sigma_{R_t} = 5$.  }

\item   Among the halo candidates for which we
have found an $\rvir$, we discard those with $N_{\rm vir}<\Npmin$,
where $N_{\rm vir}$ is the number of particles within $\rvir$.  Among
the halo candidates for which we have found a rise in spherical
density, we discard those which contain less than $N_{\rm \Rt}^{min}$
particles, where $N_{\rm
\Rt}^{min} = \Npmin$ if $\Rt > \Rvirmin$, otherwise
\begin{eqnarray}
         N_{\rm \Rt}^{\rm min} = 
         N_{\rm p}^{\rm min}{\left(\frac{\Rt}{\Rvirmin}\right)}.
\end{eqnarray}
The above constraint follows from an extrapolation of the minimum mass
halo using an isothermal profile $\rho(r) \propto 1/r^2$.

\item For halos with more than $\Npfmin$ particles, we model the 
density profile of each halo using the NFW form
(Eq~\ref{eq:nfw}) and determine the best fit $\rs$ and $\rho_s$
values, which determine $\rvir$ and $\Mvir$.  The fitting procedure
uses logarithmically spaced radial bins from max($2f_{\rm res}, 0.02
\times {\rm min}(\rvir,\Rt)$) out to $R_h$.   If any bins are empty we
decrease the number of bins by one until this is no longer the case.
If the number of bins is reduced below three we discard the halo as a
local perturbation.

The fit takes into account the Poisson error in each bin due to the
finite number of particles, and we obtain errors on the fit parameters
($\sigma_{\rs}$ and $\sigma_{\rho_s}$) using the covariance matrix in
the fit routine.  The errors on the fit parameters can be translated
easily into errors for $\rvir$ ($\sigma_{\rvir}$) and the estimated
NFW mass of each halo, $\Mvir$ ($\sigma_M$).  In some cases, the fit
does not converge.  When this occurs, we mark the halo as a non-fit.
This occurrence is rare for distinct halos, but is more common for
subhalos (see below).  This may reflect a tendency for subhalos
defined with the current merging criteria to be poorly described by an
NFW form, perhaps as a result of frequent interactions or close
neighbors.

\item We unify halos with centers that overlap by $R_{\rm combine}$.  
For a given pair of halos with virial radii $R_{\rm vir,1}$, $R_{\rm
vir,2}$, we define this combination radius to be $R_{\rm combine}$ =
min($R_{\rm vir,1}/2$, $R_{\rm vir,1}/2$).  If either of the halos
does not have a fit $\rvir$, we use the halo radius $R_h$ in place of
$\Rvir$.  Our criterion is met if two (or more) halo centers are
within $R_{\rm combine}$ of each other while at the same time having
velocities which allow them to be bound to the common system.  If such
a case occurs, then along with the individual halo NFW fits, we fit
another NFW profile about the common center of mass of the two
combined halos and decide whether the candidate-united-halos are
bound/unbound to the common NFW fit using the radial escape velocity
determined using the common NFW profile (see below).  If both halos
are bound we combine the two halos into one, and keep the common fit
for the characteristic parameters.  If at least one is not bound, we
do not combine the halos.

\item For each halo, we remove all unbound particles 
before we  obtain the final fits.  We  loop over  all particles within
the halo and declare a particle at a distance $r$ from the center of a
halo to  be unbound if   its velocity relative to   the center of mass
velocity  of  the halo    obeys  $ v   > \sqrt{2\left|\Phi(r)\right|  \,}$.
For halos that have fits, we use    
the  radial  potential   for an NFW density
profile\footnote[7]{Note that this  potential  is {\it not}
necessarily  the   physical   gravitational potential  at    the  halo
location. For a subhalo, for example, the host background potential is
{\it not} included. }:
\begin{eqnarray}
   \Phi_{\sss \rm NFW}(r) = 
  -4\pi G \rho_{\rm s} {R_{\rm s}}^2\left[ \frac{\log(1 + x)}{x}\right];
\end{eqnarray}
otherwise, we use the radial potential for a singular isothermal
sphere with the same mass.
 
After removal, we construct a new density profile (and find new NFW
fit parameters if $\Np \ge \Npfmin$).  The procedure is repeated until
the number of unbound particles becomes $< 1 \%$ of the bound
particles or until the total number of particles within the halo falls
below $\Npmin$.

\item 
For each halo in the final catalog, we determine its NFW fit
if $\Np \ge \Npfmin$, and record its fit parameters and their errors.
We also measure and record its spin parameter,
$\lambda$, and the maximum of its circular velocity curve, $V_{\rm max}$.

\end{enumerate}

As described above, the halo catalog we have developed includes an
arbitrary number of levels of substructure within halos.  The full
catalog with substructure should then include all halos directly
around galaxies, above the relevant mass resolution, and thus is
useful for a number of direct comparisons with observations.
However, for many purposes, a halo catalog including only ``distinct''
halos, i.e., halos which are not subhalos of any larger halo, is sufficient
and introduces significantly less complication. For this reason we 
have culled the full catalog into a smaller catalog that does not include
subhalos within halos; this is the catalog analyzed here.

Since there are multiple levels of substructure, the details depend
slightly on the algorithm chosen.  We take the maximum circular
velocity to be the most reliable measure of the halo's size, since it
is a measured quantity and doesn't rely on a fit, and can be defined
equivalently for all halos.  All halos whose centers lie within the
virial radius of a larger halo are then designated as subhalos.  Note
that a halo that lies within the virial radius of subhalos is only
removed if it itself is classified as a subhalo of a distinct halo.

\section{Correcting masses}
\label{sec:corrmass}
Our procedure of fitting density profiles is intended to give the best
estimate possible of a halo's virial mass; however, it is subject to
large errors when there are a small number of particles or especially
when the halo is undergoing merging or disruption and is far from
being a relaxed, spherical object.  These uncertainties are taken into
account in the fit errors, but for many purposes it is essential to
have the best estimate possible of the halo's mass at each output time.
Especially for consideration of the evolution of individual halo mass
trajectories it would be useful to eliminate large jumps in the
trajectories which are due to the above-mentioned irregularities and
not to real changes in a halo's mass.  In order to do this, for each
halo we compare three masses: $\Mvir$, as measured from the NFW fit,
$\Mh$, the measured mass within $R_h$, and $\Mtraj$, which designates
the mass interpolated between the most massive progenitor in the
previous output time and the offspring halo in the subsequent output time
(assuming they both exist).  In most cases, $M=\Mvir$ (if $\Mvir$
exists, otherwise $M=\Mh$); it is only changed if this mass seems
clearly inconsistent with the other mass estimates and does not seem
reasonable.  For most halos, the error on $\Mvir$ is small and $\Mvir
\simeq \Mh$; in these cases $M$ always equals $\Mvir$.  However, if
one of these is not the case, $\Mvir$ is used if it is close to
$\Mtraj$ and otherwise $\Mh$ is used if it is close to $\Mtraj$.  If
neither seems consistent with the halo's trajectory, we use $M={\rm
median}(\Mvir, \Mh, \Mtraj)$.  The details of the procedure are
slightly more complicated, depending on the error on $\Mvir$, and we
direct the reader to Wechsler (2001) for a complete description.

\section{Generating Merger Trees}
\label{sec:epsapp}
For completeness, we outline here the fundamental aspects of EPS
and our method for generating merger trees.
\citet[hereafter LC93]{lacey:93} introduced a method for calculating the 
probability that a halo of mass M accretes a given mass in a given
time.
Let $S(M)
\equiv \sigma^2(M)$ be the linear density variance on the mass scale
$M$ and $w(t) \equiv \delta_c(t)$ be the linear density for collapsing
structures at time $t$ (see, e.g., \citealt{white:96}).  Given a halo
of mass $M$ at some time $t$ , the probability that it accretes a mass
$\Delta M$ in a time $\Delta t$ is then
\beq
P(\Delta S, \Delta w)d\Delta S = \frac{1}{\sqrt{2 \pi}}
  \frac{\Delta w}{(\Delta S)^{3/2}}\exp\left[{-\frac{(\Delta w)^2}{2 \Delta S}}
\right] d\Delta S,
\label{eq:eps}
\eeq
where $\Delta S = S(M) - S(M + \Delta M)$ and $\Delta w = w(t) - w(t_1
+ \Delta t)$.  In order to generate halo merging trees, one must
implement this formula iteratively, with some algorithm for choosing
progenitors.  However, no method has been proposed that simultaneously
matches the the conditional mass function and progenitor distribution
of specified by Eq. \ref{eq:eps} exactly.  We use the scheme suggested
by \citet[SK99]{sk:99}, which enforces mass conservation exactly but
only reproduces the progenitor distribution of EPS approximately (for
alternative techniques see \citealt{kauf:93} and LC93).  In order to
keep the trees finite, a minimum progenitor mass, $M_m$ is defined.
Halo mass growth with $\Delta M < M_m$ is treated as diffuse
accretion.  A key ingredient of this technique is that the timestep
must be chosen such that $\Delta w \lsim
\sqrt{M_m dS/dM}$ in order to reproduce the expected conditional mass 
functions of extended Press-Schechter. Stepping back in time, the tree
then provides a list of progenitors and their masses.  In order to
better match the analytic prediction of the progenitor distribution,
we apply an addition fix to the method of SK99, suggested by
\citet{bkw:00}, which constrains the number of progenitors at any
timestep to be close to the mean for that mass and redshift.

n

\end{document}